\documentclass[manuscript, screen]{acmart}
\AtBeginDocument{%
  }

\usepackage{graphicx}
\usepackage{multirow}
\usepackage{tabularx}
\usepackage{tabularray}
\usepackage{graphicx}
\usepackage{lscape}
\usepackage{mfirstuc}
\usepackage{pifont}
\usepackage{tablefootnote}

\usepackage{enumitem}
\setlist{nosep}

\newcommand{\jp}[1]{\textcolor{black}{#1}}
\newcommand{\ts}[1]{\textcolor{black}{#1}}

\newcommand{\cmark}{\ding{51}}
\newcommand{\xmark}{\ding{55}}
\newcommand{\qmark}{?}
\newcommand{\tilemark}{$\sim$}
\newcommand{\dollarmark}{\textdollar}

\newcommand{\notePercentageRounding}{(percentages may not sum to 100\% due to rounding)}

\newcommand{\dateOfLiteratureSearch}{\ts{29th of June 2026}}

\newcommand{\totalNumberOfPapersFromPrimarySearchResult}{\ts{246}}
\newcommand{\totalNumberOfRelevantPapersFromPrimarySearchResult}{40}
\newcommand{\totalNumberOfUniqueRelevantPapersFromPrimarySearchResult}{18}
\newcommand{\totalNumberOfUniquePapersFromAllRounds}{\ts{1397}}

\newcommand{\totalNumberOfUniqueRelevantPapersFromAllRounds}{42}
\newcommand{\totalNumberOfUniqueRelevantPapersInFinalCorpus}{\ts{46}}

\newcommand{\totalNumberOfPapersSuggestedByAuthorsViaEmail}{\ts{4}}
\newcommand{\totalNumberOfPapersSuggestedByAuthorsViaEmailText}{\ts{four}}

\newcommand{\totalNumberOfDistinctTransplantationApproaches}{\ts{27}}

\newcommand{\numberOfPapersNameSoftwareForTransplantationPurpose}{\ts{40}}
\newcommand{\numberOfPapersDiscussSoftwareFeature}{31}
\newcommand{\percentageOfPapersDiscussSoftwareFeature}{\ts{78\%}}
\newcommand{\numberOfPapersDiscussTestCase}{\ts{9}}
\newcommand{\percentageOfPapersDiscussTestCase}{\ts{23\%}}

\newcommand{\numberOfBenchmarksTotal}{\ts{642}}
\newcommand{\numberOfBenchmarksNoLicenseProvided}{158}

\newcommand{\numberOfBenchmarksRetrievalRequiresToolExecution}{\ts{125}}

\newcommand{\percentageOfBenchmarksOpenSource}{\ts{16\%}}
\newcommand{\numberOfBenchmarksSourceCodeIsUnavailable}{\ts{33}}

\newcommand{\numberOfBenchmarksArtificialSoftware}{10}

\newcommand{\numberOfAvailableBenchmarksTotal}{\ts{203}}
\newcommand{\percentageOfAvailableBenchmarksTotal}{\ts{32\%}}
\newcommand{\numberOfConditionallyAvailableBenchmarksTotal}{\ts{283}}
\newcommand{\percentageOfConditionallyAvailableBenchmarksTotal}{\ts{44\%}}
\newcommand{\numberOfUnavailableBenchmarksTotal}{\ts{156}}
\newcommand{\percentageOfUnavailableBenchmarksTotal}{\ts{24\%}}

\newcommand{\numberOfBenchmarksUsedAcrossDifferentStudies}{\ts{29}}
\newcommand{\numberOfStudiesUsingRepeatedBenchmarks}{\ts{12}}
\newcommand{\numberOfBenchmarksUsedAsBothDonorAndHost}{\ts{107}}

\newcommand{\percentageOfOrganLanguageJava}{\ts{35\%}}

\newcommand{\percentageOfOrganLanguageC}{\ts{15\%}}

\newcommand{\percentageOfOrganLanguageNotMentioned}{\ts{20\%}}

\newcommand{\percentageOfBenchmarkLanguageJava}{\ts{21\%}}

\newcommand{\percentageOfBenchmarkLanguageC}{\ts{15\%}}

\newcommand{\percentageOfBenchmarkLanguageNotMentioned}{\ts{30\%}}

\newcommand{\percentageOfFeatureReuseApproaches}{\ts{74\%}}
\newcommand{\percentageOfTestReuseApproaches}{\ts{26\%}}
\newcommand{\percentageOfApproachesUsingManualOrganLocation}{88\%}

\newcommand{\numberOfApproachesUsingInternalValidationStep}{\ts{10}}
\newcommand{\percentageOfApproachesUsingInternalValidationStep}{\ts{37\%}}
\newcommand{\numberOfApproachesUsingInternalValidationStepViaTest}{\ts{nine}}
\newcommand{\numberOfApproachesUsingInternalValidationStepViaOtherThanTest}{\ts{two}}
\newcommand{\numberOfApproachesUsingExternalValidationStep}{\ts{13}}
\newcommand{\percentageOfApproachesUsingExternalValidationStep}{\ts{48\%}}
\newcommand{\numberOfApproachesManualExternalValidationStep}{\ts{eight}}
\newcommand{\percentageOfApproachesManualExternalValidationStep}{\ts{30\%}}
\newcommand{\numberOfApproachesNoMentioningOfValidationStep}{four}
\newcommand{\percentageOfApproachesNoMentioningOfValidationStep}{\ts{15\%}}

\newcommand{\numberOfApproachesRequireUserInterventionDuringAdaptation}{two}
\newcommand{\percentageOfApproahcesRequireUserInterventionDuringAdaptation}{\ts{7\%}}

\newcommand{\percentageOfApproahcesWithAutomationDuringAdaptation}{\ts{93\%}}

\newcommand{\numberOfStudyThatInvolveHumanParticipants}{seven}

\newcommand{\percentageOfProfessionalSoftwareEngineersInUserStudy}{39\%}

\newcommand{\numberOfApproachesLabelAllReuseDecision}{one}
\newcommand{\percentageOfApproachesLabelAllReuseDecision}{4\%}

\newcommand{\numberOfApproachesUserProvidesSearchQuery}{5}
\newcommand{\percentageOfApproachesUserProvidesSearchQuery}{\ts{19\%}}

\newcommand{\numberOfApproachesNotRequireOrganLocation}{three}
\newcommand{\percentageOfApproachesNotRequireOrganLocation}{\ts{11\%}}
\newcommand{\numberOfApproachesUserProvidesOrganLocation}{\ts{18}}
\newcommand{\percentageOfApproachesUserProvidesOrganLocation}{\ts{67\%}}

\newcommand{\percentageOfApproachesUsingCodeSearchEngineForOrganExtraction}{\ts{19\%}}

\newcommand{\percentageOfApproachesUsingStaticAnalysisForOrganExtraction}{\ts{30\%}}

\newcommand{\percentageOfApproachesUsingDynamicAnalysisForOrganExtraction}{\ts{30\%}}

\newcommand{\percentageOfApproachesUsingStaticAndDynamicAnalysisForOrganExtraction}{4\%}

\newcommand{\percentageOfApproachesNoOrganExtraction}{\ts{19\%}}

\newcommand{\percentageOfApproachesUsingProgramSlicingForOrganExtraction}{\ts{15\%}}

\newcommand{\percentageOfApproachesUsingStaticAnalysisForOrganAdaptation}{\ts{48\%}}

\newcommand{\percentageOfApproachesUsingDynamicAnalysisForOrganAdaptation}{\ts{48\%}}

\newcommand{\percentageOfApproachesUsingStaticAndDynamicAnalysisForOrganAdaptation}{4\%}
\newcommand{\numberOfApproachesUsingGPForOrganAdaptation}{five}
\newcommand{\percentageOfApproachesUsingGPForOrganAdaptation}{\ts{19\%}}

\newcommand{\percentageOfApproachesUsingGUIEventMappingForOrganAdaptation}{\ts{22\%}}

\newcommand{\numberOfApproachesUserProvidesInsertionLocation}{12}
\newcommand{\percentageOfApproachesUserProvidesInsertionLocation}{\ts{44\%}}
\newcommand{\numberOfApproachesRandomlyInsertOrgan}{one}
\newcommand{\percentageOfApproachesRandomlyInsertOrgan}{4\%}
\newcommand{\numberOfApproachesDoNotMentionInsertionLocation}{\ts{14}}
\newcommand{\percentageOfApproachesDoNotMentionInsertionLocation}{\ts{52\%}}

\newcommand{\numberOfApproachesAdaptingOrganAfterInsertion}{six}
\newcommand{\percentageOfApproachesAdaptingOrganAfterInsertion}{24\%}
\newcommand{\numberOfApproachesInsertingOnlyMissingPartOfOrganIntoHost}{two}
\newcommand{\percentageOfApproachesInsertingOnlyMissingPartOfOrganIntoHost}{8\%}
\newcommand{\numberOfApproachesEnumeratePossibleInsertionLocations}{one}
\newcommand{\percentageOfApproachesEnumeratePossibleInsertionLocations}{4\%}

\newcommand{\numberOfDistinctSoftwareTransplantationTools}{\ts{26}}
\newcommand{\numberOfToolsAvailable}{\ts{13}}
\newcommand{\percentageOfToolsAvailable}{\ts{50\%}}

\newcommand{\percentageOfToolsUnavailable}{\ts{31\%}}

\newcommand{\percentageOfToolsPendingResponse}{\ts{19\%}}

\newcommand{\numberOfAvailableToolsFullyRunnable}{six}
\newcommand{\numberOfAvailableToolsPartiallyRunnable}{\ts{two}}
\newcommand{\numberOfAvailableToolsPendingResponse}{three}
\newcommand{\numberOfAvailableToolsNoLongerOperational}{one}
\newcommand{\numberOfAvailableToolsNotVerified}{one}

\usepackage[framemethod=tikz]{mdframed}
\mdfsetup{
    linewidth=1pt,
    backgroundcolor=gray!20,
    innertopmargin=3pt,
    innerbottommargin=3pt,
    innerrightmargin=3pt,
    innerleftmargin=3pt,
    skipabove=3pt,
    skipbelow=3pt,
}

\begin{document}
\title[Reusing Battle-tested Code: A Systematic Literature Review of Automated Software Transplantation]{Reusing Battle-tested Code: A Systematic Literature Review of Automated Software Transplantation}

\author{Thanatad Songpetchmongkol}
\email{thanatad.songpetchmongkol.22@ucl.ac.uk}
\orcid{0009-0007-8615-149X}
\affiliation{%
  \institution{University College London}
  \city{London}
  \country{United Kingdom}
}

\author{Earl T. Barr}
\email{e.barr@ucl.ac.uk}
\orcid{0000-0003-0771-7891}
\affiliation{%
  \institution{University College London}
  \city{London}
  \country{United Kingdom}
}

\author{Mark Harman}
\email{mark.harman@ucl.ac.uk}
\orcid{0000-0002-5864-4488}
\affiliation{%
  \institution{University College London}
  \city{London}
  \country{United Kingdom}
}

\author{Federica Sarro}
\email{f.sarro@ucl.ac.uk}
\orcid{0000-0002-9146-442X}
\affiliation{%
  \institution{University College London}
  \city{London}
  \country{United Kingdom}
}

\author{Justyna Petke}
\email{j.petke@ucl.ac.uk}
\orcid{0000-0002-7833-6044}
\affiliation{%
  \institution{University College London}
  \city{London}
  \country{United Kingdom}
}

\begin{abstract}
Software reuse remains a fundamental challenge in software engineering, often caught between the security risks of external libraries and the reliability concerns of generative AI. 
Automated software transplantation offers a compelling alternative by enabling the direct extraction and integration, into new host systems, of battle-tested logic (code that has already survived the rigors of production, 
possibly over many years of extensive in-production use). 
This paper presents a systematic literature review to evaluate the current state-of-the-art and identify the remaining barriers preventing this technology from becoming a mainstream engineering practice. 
We identified \totalNumberOfDistinctTransplantationApproaches{} distinct automated software transplantation approaches. 
Our analysis reveals that while the adaptation of code is highly automated, significant bottlenecks remain in the manual prerequisite steps of organ location and test suite augmentation. 
We conclude with a research roadmap that suggests how these manual obstacles might be overcome, including the strategic use of Large Language Models to assist in identifying and verifying high-value, production-hardened logic for transplantation.\looseness=-1

\end{abstract}


\maketitle
\section{Introduction}
\label{sec:intro}

Software transplantation is the automated process of extracting a specific behaviour --- an ``organ'' --- from a donor system and integrating it into a host system. 
First formalized in an invited keynote paper for the Working Conference on Reverse Engineering (WCRE 2013)\citet{genetic_programming_for_reverse_engineering}, 
this field addresses the fundamental challenge of software reuse: the ability to build on existing components rather than starting from scratch~\citep{software_reuse_research_status_and_future}. 
While McIlroy at the 1968 NATO Software Engineering conference envisioned a world of mass-produced, modular components~\citep{mass_produced_software_components}, the practical reality of software engineering remains divided between structured integration and ad-hoc copying.
Although transplantation was not formally defined until 2013, as this survey shows, there was work on related ideas dating back to 2008.

Today, the dominant mechanism for reuse is the software library. By leveraging pre-existing packages, developers can integrate complex functionality with minimal effort. However, libraries are not a panacea; they introduce significant \emph{supply chain vulnerabilities}, where a security flaw in a single dependency can compromise thousands of downstream systems. Furthermore, libraries are often difficult to efficiently adapt to the needs of a particular codebase, may suffer from ``bloat'' by including unnecessary features, or a suitable library may simply not exist for niche requirements.

To manage these variations more formally, Software Product Line (SPL) engineering was introduced to facilitate reuse across sets of systems that share common features~\citep{software_product_line_engineering_and_variability_management}. By managing core assets and variations (often via conditional compilation or preprocessor directives), SPLs allow for rapid development of related products. However, SPL architecture requires substantial upfront investment and is difficult to adopt when it is not  clear if a software will evolve into multiple variants~\citep{easing_the_transition_to_software_mass_customization}. Furthermore, a vast majority of reuse opportunities remain unsuitable for SPL engineering, as they involve ``one-off'' cross-project integrations or occur in domains where the lack of stability and high fragmentation make the creation of a unified platform technically and economically infeasible~\citep{minich_industrialising_2013,clements_simple_2005}.

More recently, Large Language Models (LLMs) have introduced a transformative paradigm for reuse by distilling the vast corpus of code within their training data into a generative interface. While they have revolutionised reuse, they do not completely solve the problem for three key reasons. First, the generated output is frequently characterized by inefficient~\citep{effi_code_jie_zhang} and verbose code~\citep{whats_wrong_with_your_code_generated_by_large_language_models}. Second, LLMs can produce ``subtly incorrect'' code --- logic that appears plausible but contains edge-case bugs or security vulnerabilities that are difficult for human reviewers to detect~\citep{bugs_in_llms_tambon, ticoder_subtle_bugs}. Third, while they outperform previous generative technologies, LLMs still struggle with the ``long tail'' of reuse: rare, domain-specific, or proprietary logic underrepresented in their training sets. Ultimately, LLM-powered development trades the challenge of writing code for the new, non-trivial overhead of prompt engineering and verification, a shift that is not always a net win for developer productivity~\citep{towards_understanding_the_characteristics_of_code_generation_errors_made_by_large_language_models,do_advanced_models_eliminate_prompt_engineering}.

Software transplantation seeks to bridge this gap by providing the flexibility of manual reuse with the rigour of automation, predicated on the principle that the most reliable code is that which has already survived the rigours of production. By isolating logic at the functional level, it provides a means to reuse battle-tested code without the security risks of external libraries or the architectural overhead of SPLs,
or the reliability concerns inherent in probabilistic LLM-generated code. Automated transplantation offers a verifiable alternative by reusing existing, battle-tested logic from known-good donor systems.

Despite the growth in research into software transplantation since the WCRE keynote~\citep{genetic_programming_for_reverse_engineering}, there has been no systematic literature review on the topic. This paper surveys the state-of-the-art from 2008 to 2025, identifying the current capabilities and automation levels of proposed techniques. Our contributions include:

\begin{itemize}
    \item \textbf{A Comprehensive Taxonomy:} We analyze \totalNumberOfDistinctTransplantationApproaches{} automated transplantation approaches, categorizing them by the reused artefact (e.g., features or test cases) and the underlying techniques used for adaptation, such as Genetic Programming~\cite{genetic_improvement_survey}.
    \item \textbf{Identification of Bottlenecks:} We highlight the barriers to industrial adoption. Specifically, we find that \percentageOfApproachesUsingManualOrganLocation{} of techniques still require engineers to manually locate the organ, and \percentageOfApproachesUsingExternalValidationStep{} rely on external validation steps.
    \item \textbf{Analysis of Automation Gaps:} While \percentageOfApproahcesWithAutomationDuringAdaptation{} of approaches fully automate the adaptation phase, in which the extracted code is modified to fit the host system's environment, the whole transplantation pipeline is not yet fully automated. We highlight potential of LLMs to close this gap. 
    \item \textbf{Future Research Roadmap:} We provide recommendations for moving beyond C and Java to dynamic languages like Go and JavaScript, and suggest how LLMs might be used to assist in the manual organ location and test suite augmentation obstacles we identified.
\end{itemize}

\section{Software Transplantation History} \label{sec:software_transplantation_history}

Software engineers recognize that the functionalities they require often already exist elsewhere, either in external projects or within their own historical codebases~\citep{procrustes}. Reusing source code is not merely a matter of convenience; it is a strategy to reduce time-to-market and leverage existing, well-tested APIs that have already undergone the rigors of production use~\citep{hunter}. Consequently, the history of software engineering is, in many ways, a history of attempting to formalize the transfer of functionality between systems~\citep{code_carbon_copy}.

This pursuit has traditionally been divided into two camps: preplanned and ad-hoc reuse. In the early days, research focused heavily on structured, preplanned approaches such as object-oriented inheritance and modular encapsulation~\citep{encapsulation_and_inheritance_oop}. By the turn of the millennium, this culminated in Software Product Line (SPL) engineering, which sought to manage variations across related systems through a shared core of assets~\citep{software_product_line_engineering_and_variability_management}. However, as~\citet{gilligan} argued, these structured approaches face three fundamental hurdles: \begin{enumerate} \item It is difficult to predict which components will actually require reuse in the future. \item The upfront economic investment required to make every component ``reusable'' is often prohibitive. \item Every component carries implicit assumptions about its original execution context that may not hold during reuse. \end{enumerate}

Because of these constraints, the ``gold standard'' of mass-produced, modular components envisioned in 1968 has largely been replaced by the software library. While libraries offer ease of integration, they create a ``black box'' dependency that introduces supply chain vulnerabilities and significant code bloat. When a suitable library does not exist, or when an SPL is too rigid to accommodate a ``one-off'' cross-project need, engineers are left with three sub-optimal choices: reimplementing from scratch, performing extensive manual refactoring, or resorting to ad-hoc ``copy-paste'' reuse.

The latter, while common, is notoriously error-prone. Bugs frequently arise when copied code is modified to fit a new environment, leading to inconsistencies in identifier names or logic errors~\citep{smartPaste}. While Large Language Models (LLMs) have recently emerged as a tool for synthesizing these code snippets on demand, they largely replace the task of coding with the task of verification. Because LLMs are probabilistic rather than deterministic, they are currently unproven for generating the large-scale, robust features required in professional systems. They lack the ``battle-tested'' pedigree of code that has already functioned successfully in a live donor system.

The software transplantation proposal by~\citet{genetic_programming_for_reverse_engineering} was designed to solve this  dilemma by providing the rigour of automation to the flexibility of ad-hoc reuse. 

In this context, we draw a biological analogy: lines of code correspond to cells or tissues which form a functional ``organ''. Just as a biological organ relies on the body's systemic interactions, a software organ is typically a unit of code that must be adapted to its new environment. We thus define an \emph{organ} as follows:

\textbf{Definition 1 (Organ):} An \textit{organ} is a collection of code statements that together implement a specific, stakeholder-visible function. This function must be mappable to the requirements of the software system.

\citet{genetic_programming_for_reverse_engineering} proposed the LATIIV$^R$ approach, outlining seven steps essential for a successful ``organ'' transfer: 

 \begin{enumerate} \item \textit{Localise}: Locate the desired organ in the donor. \item \textit{Abstract}: Create an abstraction that retains the control and data flow of the organ. \item \textit{Target}: Find a target location in the host where the organ could be integrated. \item \textit{Interface}: Construct an interface allowing the host to access the organ's functionality. \item \textit{Insert}: Transplant the organ into the target location. \item \textit{Validate}: Confirm the host has gained the new functionality without regression faults. \item \textit{Repeat}: Iterate until the transplantation succeeds. \end{enumerate}

\section{Survey Methodology} \label{sec:survey_method}
This section outlines the methodology we followed for the systematic literature review, including the search engines and the scope employed. We present the number of relevant publications found at each stage of the snowballing process.

We performed a preliminary search to ensure that relevant keywords were found. 
Initially, we conducted a search with the `software transplantation' keyword. 
From the query result, we found that some publications refer to this domain as `code transplantation'. 
Consequently, we added code transplantation to the list of final keywords. 
We also tried `feature migration' and `code migration' as these often appear in the software product line literature. 
However, we found the results not particularly relevant. 
Moreover, we came across the work from~\citet{prodScalpel}, which claims to be the first to realise software transplantation in the software product line domain. 
Consequently, we did not include these as part of the final keyword set. 
Thus, our final keyword set includes \textit{`software transplant'} and \textit{`code transplant'} and their extensions (i.e., software/code transplants, software/code transplanted, software/code transplantation and software/code transplanting).

Based on the keywords identified in our preliminary search, we conducted a systematic literature review~\cite{guideline_for_performing_systematic_literature_review}. We performed keyword searches across four databases storing research articles on software engineering: IEEEXplore~\cite{ieeexplore}, ACM Digital Library~\cite{acm_digital_library}, DBLP~\cite{dblp}, and ScienceDirect~\cite{sciencedirect}. 

For all search engines, we used `full text' and `metadata' as the filter except for DBLP where we used the default filter option (i.e., `combined dblp search'). 
All keyword searches were conducted on the \dateOfLiteratureSearch{}.
Table~\ref{table:primary_search_result_small_version} shows the result of this primary search, which returned \totalNumberOfPapersFromPrimarySearchResult{} papers overall.

\begin{table}
    \centering
    \caption{Primary search results. All keyword searches were conducted on \dateOfLiteratureSearch{}.}
    \begin{tabular}{cccc}
        \hline
        Keyword & Search Engine & Papers & Relevant \\
         &  & Found & Papers\\
        \hline
         & IEEE & \ts{61} & 9\\
        Code & ACM & \ts{43} & 7\\
        Transplant* & ScienceDirect & 10 & 0\\
         & DBLP & 5 & 2 \\
        \hline
         & IEEE &\ts{75} & 7 \\
        Software & ACM & 31 & 6\\
        Transplant* & ScienceDirect & 10 & 5 \\
         & DBLP & \ts{11} & 4 \\
        \hline
        \multicolumn{2}{l}{Total number of papers}  & \totalNumberOfPapersFromPrimarySearchResult{} & \totalNumberOfRelevantPapersFromPrimarySearchResult{} \\
        \multicolumn{2}{l}{Distinct number of papers} &  & \totalNumberOfUniqueRelevantPapersFromPrimarySearchResult{} \\
        \hline
    \end{tabular}
    \label{table:primary_search_result_small_version}
\end{table}

At this stage, we started filtering papers based on relevance. 
We only include journal, conference, workshop papers, books and PhD theses in our review. 
This is because such publications have usually been peer-reviewed.
We consider a paper to be relevant if it (1) is in the computer science domain, (2) presents approaches, empirical work, surveys, and future directions for automated software transplantation, and (3) is written in English. 
To be more specific, we consider a technique relevant to the automated software transplantation domain if it satisfies the following criteria:
\begin{enumerate}
    \item An approach consists of a donor, an organ and a host.
    \item An organ performs a specific function of interest. 
    \item Organ transplantation can take place from:
    \begin{enumerate}
        \item one software to another, or
        \item within the same software.
    \end{enumerate}
    \item The proposed technique performs an automated organ adaptation procedure in which an organ is adapted to the host's lexical context at the implantation location.
\end{enumerate}

There are two main reasons why we include automated organ adaptation as one of the criteria. 
First, many publications do not necessarily move software features. 
Many publications simply present a new method of efficiently searching for code snippets that could be reused~\cite{parseweb_a_programmar_assistant} or act as a guide to confirm that the engineers correctly implemented the desired functionality~\cite{example_centric_programming_integrating_web_search, approximate_structural_context_matching}. 
Second, we are particularly interested in the automated aspects of the software transplantation techniques. 
If the proposed technique merely injects the code into the host in the original form (i.e., as-is), then the responsibility for adapting the code (e.g., renaming variables) falls entirely on the engineers~\cite{NLP2Code}. 
This approach does not provide any added value compared to doing the work manually. 
Therefore, we consider such papers to be irrelevant to our survey.

\jp{Although S6~\cite{semantics_based_code_search} performs a form of code adaptation, we exclude it because its adaptation is not performed for a specific host. 
S6 is a method of code search that additionally provides a set of transformations, which the user can choose to perform on the extracted code. 
Consequently, if the user decides to reuse the resulting code in a host system, they may still need to manually adapt it to the host's context.}

\jp{Moreover, we are interested in transplantation of an organ performing a specific function of interest. 
CodePhage~\cite{Sidiroglou-Douskos15:codephage}, for instance, uses transplantation as means of code repair.
CodePhage identifies an organ similar to existing functionality except that it passes tests that fail on the faulty code. 
Via program analysis only code fragments are transplanted that fix the issue rather than a full function itself. 
Authors of CodePhage have also released other work on transplantation which we include~\cite{code_carbon_copy,program_fracture_and_recombination_for_efficient_automatic_code_reuse,horizontal_code_transfer_via_program_fracture_and_recombination} as it transfers whole functionality rather than a patch between the original and repaired code.
}

Based on our criteria, we identified \totalNumberOfUniqueRelevantPapersFromPrimarySearchResult{} relevant papers from the primary search results. 
Then, we started the snowballing process, where we manually went through each reference from the papers that are in our scope. 
This process ended when no new relevant papers were found. 
At the end of the process, we have evaluated \totalNumberOfUniquePapersFromAllRounds{} unique papers for relevance and found \totalNumberOfUniqueRelevantPapersFromAllRounds{} relevant papers (shown in Tables~\ref{tab:papers_summary_rq1_to_rq2}, ~\ref{tab:papers_summary_rq3_to_rq5} 
and~\ref{tab:papers_summary_rq6_to_rq8}) from three rounds of snowballing (see Table~\ref{tab:snowballing_result}).
Approximately 10\% of the papers (159) were then selected for evaluation by one of the other authors of the survey.
There was disagreement on 6 of those, which were then discussed for inclusion. In the end, we kept 2 of 6 in the literature review. 

\jp{Overall, \totalNumberOfUniqueRelevantPapersFromAllRounds{} were identified. 
To ensure we did not miss any relevant literature, we emailed contact authors of all the \totalNumberOfUniqueRelevantPapersFromAllRounds{} relevant papers found. 
Via this route we identified \totalNumberOfPapersSuggestedByAuthorsViaEmailText{} new publications within our search range, which we include. 
Our final corpus thus consists of \totalNumberOfUniqueRelevantPapersInFinalCorpus{} unique relevant publications.
}

\begin{table}
    \centering
    \caption{Number of papers found in each round of snowballing as well as the total number of unique, relevant papers found on the topic of software transplantation.}
    \begin{tabular}{cccc}
        \hline
        Round & Processed Papers & New Unique & New Relevant \\
        \hline
        0 & \totalNumberOfPapersFromPrimarySearchResult{} & \ts{174} & \totalNumberOfUniqueRelevantPapersFromPrimarySearchResult{}\\
        1 & 1,166 & 891 & 14\\
        2 & 289 & 183 & 10\\
        3 & 325 & 149 & 0\\
        \hline
        Total & \ts{2,026} & \totalNumberOfUniquePapersFromAllRounds{} & \totalNumberOfUniqueRelevantPapersFromAllRounds{} \\
        \hline
    \end{tabular}
    \label{tab:snowballing_result}
\end{table}

\begin{table}
    \centering
    \caption{Summary of papers from literature review on software transplantation.}
    \begin{tabular}{lr}
        \hline
        Search Step & Total Unique Relevant Paper Found\\
        \hline
        Primary & \totalNumberOfUniqueRelevantPapersFromPrimarySearchResult{}\\
        Snowballing & 24\\
        Suggestions from authors & \totalNumberOfPapersSuggestedByAuthorsViaEmail{}\\
        \hline
        Papers on software transplantation & \totalNumberOfUniqueRelevantPapersInFinalCorpus{}\\
        \hline
    \end{tabular}
    \label{tab:summary_of_paper_from_literature_review_of_software_transplantation}
\end{table}

\begin{figure}
    \centering
    \includegraphics[width=0.60\linewidth]{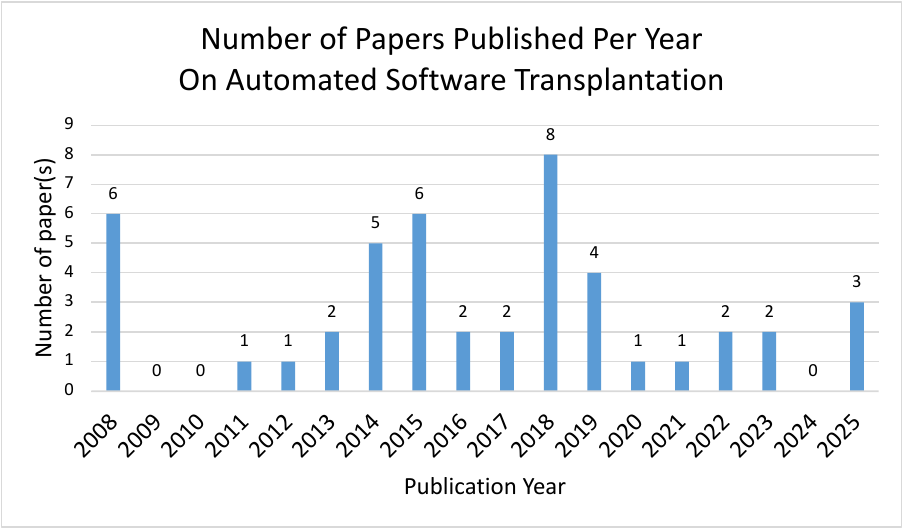}
    \caption{Publication distribution of \totalNumberOfUniqueRelevantPapersInFinalCorpus{} papers found from the systematic literature review on Software Transplantation.}
    \Description{}
    \label{fig:literature_review_publication_per_years}
\end{figure}

Figure~\ref{fig:literature_review_publication_per_years} shows the number of publications found per year. 
Overall, the number of publications remains relatively low, with a maximum of only eight publications in a given year.
In some years, no publications were recorded at all. 
Before the keynote speech by Harman in 2013, research was centred around copy-and-paste (i.e., focusing on the small-scale reuse of source code), pragmatic software reuse, and web user interface reuse.
Following the keynote, there was a noticeable increase in interest in the software transplantation domain, with a peak in publications between 2013 and 2018. 
However, this trend did not persist and interest appears to have decreased afterwards. 
This decline in research activity could help explain why, despite the domain being formally introduced over a decade ago, only \totalNumberOfUniqueRelevantPapersInFinalCorpus{} publications were found \ts{through our search and authors' suggestions.}

\begin{figure}
    \centering
    \includegraphics[width=0.6\linewidth]{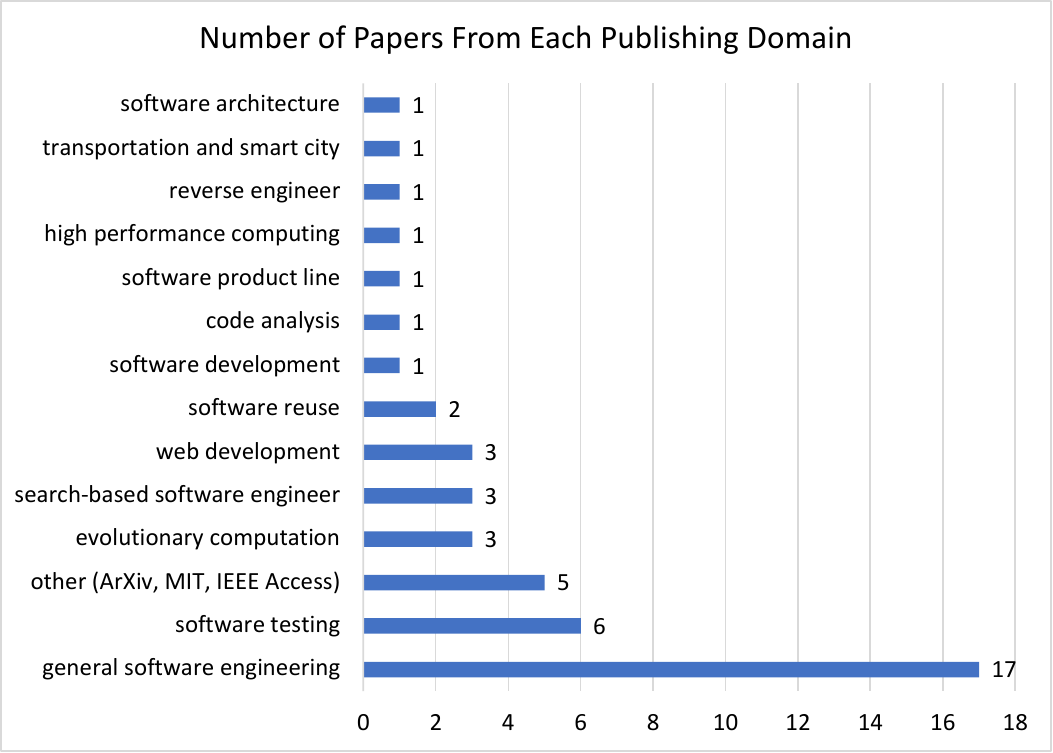}
    \caption{The number of papers from each publishing venue type. The majority of papers were published in a `general software engineering' venue with \ts{17} publications, which includes ICSE, ASE, FSE, and others. This is followed by software testing (e.g., ISSTA, ICST), evolutionary computation (e.g., GECCO), and search-based software engineering (e.g., SSBSE).}
    \Description{}
    \label{fig:number_of_paper_per_publishing_domain}
\end{figure}

Figure \ref{fig:number_of_paper_per_publishing_domain} presents the distribution of publications across different publishing domains, including general software engineering, evolutionary computation, software reuse, and others. 
It is evident that the majority of work is published in general software engineering venues.
Additionally, there are \ts{six} papers published in software testing venues and three papers each published in evolutionary computation and search-based software engineering.
This indicates that the researchers in these fields recognise the potential and benefit of employing software transplantation as a method of engineering software.
It also suggests that some proposed techniques rely on heuristic algorithms as their core methodology while framing transplantation as an optimisation problem. 
Furthermore, the importance of testing is highlighted to ensure that the host can exercise the new functionality and no regression fault occurs.

\section{Research Questions}
The main goal of this systematic literature review is to understand the current state-of-the-art in software transplantation. 
Given that software transplantation promises great potential to optimise the software development process, we want to see to what extent this has been utilised in the industrial setting. 
Given our goals, we present our research questions as follows:

\textbf{RQ1: How do various transplantation approaches locate an organ?}
Before the transplantation can start, we need to know where the organ is located in the donor. 
Only then, extraction and implantation of the organ into the host can take place. 
Therefore, we look into the approaches that locate the function of interest.

\textbf{RQ2: What types of methods have been proposed for the extraction of the identified organ from the donor?}
After the organ has been located, it needs to be extracted from the donor.
This is a crucial step, as the extracted code needs to contain all information about the organ, including all its dependencies, and interface that would make the organ to be executable.
Hence, we look at the methods for performing the organ extraction task.

\textbf{RQ3: What types of methods have been proposed for transplanting the identified organ into the host?}
To answer this, we look at the methods for performing the organ adaptation task. 
Specifically, we are interested in the procedure of how one would inject an organ into the host, adapting it to the host's environment.

\textbf{RQ4: How have the results of software transplantation been validated?}
For transplantation to be considered successful, the host should be able to exercise the new functionality of the organ. 
This new functionality should work correctly. 
Concurrently, the host's existing functionalities should be preserved. 
To achieve this, we report on methods that check the above.

\textbf{RQ5: To what extent has software transplantation been automated?}
We examine each step of the transplantation process for each proposed approach to establish whether it requires human interventaion or not.

\textbf{RQ6: What type of software has been used in the software transplantation domain?}
We want to establish what types of organs have been successfully transplanted, e.g., features or tests, and their characteristics, such as size and programing language.
We also report on re-usability of existing artefacts from software transplantation studies for future benchmarking pruposes.

\textbf{RQ7: To what extent has the software transplantation been adopted in an industrial setting?}
Although software transplantation seems promising in terms of increasing software development efficiency and promoting a higher degree of automation, it is not clear to what extent this has been adopted in an industrial setting. 
We answer this question by looking at previous publications to identify how many proposed approaches have been tried in the industry and/or used industrial software.

\textbf{RQ8: How many software transplantation tools from the literature are still available and runnable today?}
We aim to assess the extent to which existing tools remain accessible and executable, either for researchers wishing to extend prior work or for software engineers interested in exploring and applying them in real-world settings. 
To answer this question, we collected the source code of the tools and attempted to run them.

\section{Results}

This section presents the results from the systematic literature review and answers each research question in turn. 
Tables~\ref{tab:papers_summary_rq1_to_rq2}, ~\ref{tab:papers_summary_rq3_to_rq5} and~\ref{tab:papers_summary_rq6_to_rq8} list the papers used to answer each corresponding research question.

\begin{table*}
    \caption{Publications on automated software transplantation, with summary of answers to our research questions (RQ1–RQ2). Tool names added, if provided in a given publication.}
    \label{tab:papers_summary_rq1_to_rq2}
    \centering
    \begin{tabular}{lll}
        \hline
        Tool Name / Paper & RQ1: location & RQ2:  extraction \\
        \hline
        Code Conjurer \cite{code_conjurer}& search query & code search engine \\
        Jigsaw \cite{jigsaw_fullPaper} & user-provided code & no extraction \\
        Jigsaw \cite{jigsaw_2pages} & user-provided code & no extraction \\
        \citet{semi_automating_pragmatic_reuse_tasks} & label all components & static analysis \\
        Procrustes \cite{procrustes} & label all components & static analysis \\
        Gilligan \cite{pragmatic_software_reuse_PhD} & label all components & static analysis \\
        \citet{reusing_web_application_user-interface_controls} & user-provided code &  static and dynamic analysis \\
        Gilligan \cite{gilligan} & label all components & static analysis \\
        \citet{genetic_programming_for_reverse_engineering} & - & - \\
        \citet{towards_automatic_client-side_feature_reuse} & user-provided code & dynamic analysis \\
        \citet{SBSE_for_SPL} & - & - \\
        \citet{justyna_specialise_minisat} & not required & static analysis \\
        Grow and Graft \cite{babel_pidgin} & not required &  no extraction\\
        \(\mu Scalpel\) \cite{muScalpel} & label the start of the organ &  static analysis\\
        \citet{towards_specifying_pragmatic_software_reuse} & - &  -\\
        \(\mu Scalpel\) \cite{muScalpel_kate_ide} & label the start of the organ &  static analysis\\
        Grow and Serve \cite{grow_django_citation} & not required &  no extraction\\
        \citet{program_fracture_and_recombination_for_efficient_automatic_code_reuse} & search query & code search engine \\
        \citet{horizontal_code_transfer_via_program_fracture_and_recombination} & search query & code search engine \\
        Hunter \cite{hunter} & search query & code search engine\\
        Hunter \cite{hunter_fullPaper} & search query & code search engine \\
        CodeCarbonCopy \cite{code_carbon_copy} & user-provided code &  dynamic analysis \\
        \citet{smartPaste} & user-provided code & no extraction \\
        \citet{a_study_on_tree-based_code_extraction} & user-provided code & static analysis\\
        \citet{an_initial_step_toward_organ_transplantation} & - & - \\
        \citet{genetic_improvement_survey} & - & - \\
        Splicer \cite{program_splicing} & search query & code search engine \\
        \citet{justyna_specialise_downstream} & not required & static analysis \\
        \citet{a_study_on_code_transplantation_technique} & user-provided code & static analysis \\
        AppTestMigrator \cite{automated_test_migration_for_mobile_apps} & user-provided code &  dynamic analysis \\
        GUITestMigrator \cite{test_migration_for_efficient_large-scale_assessment_of_mobile_app_coding_assignments} & user-provided code & dynamic analysis \\
        \citet{dissection_on_java_organs_in_github_repositories} & - & - \\
        CraftDroid \cite{test_transfer_across_mobile_apps_through_semantic_mapping} & user-provided code & dynamic analysis \\
        TestMig \cite{testmig_migrating_gui_test_cases_from_ios_to_android} & user-provided code & dynamic analysis \\
        AppTestMigrator \cite{test_migration_between_mobile_apps_with_similar_functionality} & user-provided code & dynamic analysis \\
        NLP2TestableCode \cite{NLP2TestableCode} & search query & code search engine \\
        \citet{semi-automated_test-case_propagation_in_fork_ecosystem} & - & - \\
        \citet{test_transplantation_through_dynamic_test_slicing} & user-specified target state & dynamic analysis \\
        AdaptivePaste \cite{adaptivePaste} & user-provided code & no extraction \\
        \ts{TransDroid \cite{gui_test_transfer_from_web_to_android}} & \ts{user-provided code} & \ts{dynamic analysis}\\
        \citet{software_testing_research_challenges_an_industrial_perspective} & - & - \\
        ProdScalpel \cite{prodScalpel} & label the start of the organ &  static analysis \\
        \citet{syntax_preserving_program_slicing} & user-specified slicing criterion & static analysis \\
        \ts{LLMigrate \cite{automated_test_transfer_across_android_apps_using_large_language_models}} & \ts{user-provided code} & \ts{dynamic analysis} \\
        \hline
    \end{tabular}
\end{table*}

\begin{table*}
    \caption{Publications on automated software transplantation, with summary of answers to our research questions (RQ3-RQ5). Tool names added, if provided in a given publication.}
    \label{tab:papers_summary_rq3_to_rq5}
    \centering
    \begin{tabular}{llll}
        \hline
        Tool Name / Paper &  RQ3: adaptation &RQ4: validation & RQ5: automation \\
        \hline
        Code Conjurer \cite{code_conjurer}&  static analysis&integrated 	& \cmark{} \\
        Jigsaw \cite{jigsaw_fullPaper} &   static analysis&external	&  \cmark{}\\
        Jigsaw \cite{jigsaw_2pages} &   static analysis&external	&  \cmark{}\\
        \citet{semi_automating_pragmatic_reuse_tasks} &   static analysis&external	&  \cmark{}\\
        Procrustes \cite{procrustes} &   static analysis&external	&  \cmark{}\\
        Gilligan \cite{pragmatic_software_reuse_PhD} &   static analysis&external	&  \cmark{}\\
        \citet{reusing_web_application_user-interface_controls} &   static analysis&not mentioned	&  \cmark{}\\
        Gilligan \cite{gilligan} &   static analysis&external	&  \cmark{}\\
        \citet{genetic_programming_for_reverse_engineering} &   -&-	&  -\\
        \citet{towards_automatic_client-side_feature_reuse} &   static analysis&external	&  \cmark{}\\
        \citet{SBSE_for_SPL} &   -&-	&  - \\
        \citet{justyna_specialise_minisat} &   dynamic analysis&integrated	&  \cmark{}\\
        Grow and Graft \cite{babel_pidgin} &   dynamic analysis&integrated	&  \cmark{}\\
        \(\mu Scalpel\) \cite{muScalpel} &   dynamic analysis&integrated	&  \cmark{}\\
        \citet{towards_specifying_pragmatic_software_reuse} &   -&-	&  - \\
        \(\mu Scalpel\) \cite{muScalpel_kate_ide} &   dynamic analysis&integrated	&  \cmark{}\\
        Grow and Serve \cite{grow_django_citation} &   dynamic analysis&integrated	&  \cmark{}\\
        \citet{program_fracture_and_recombination_for_efficient_automatic_code_reuse} &   dynamic analysis&integrated	&  \cmark{}\\
        \citet{horizontal_code_transfer_via_program_fracture_and_recombination} &   dynamic analysis&integrated	&  \cmark{} \\
        Hunter \cite{hunter} &   static analysis&integrated	&  \cmark{} \\
        Hunter \cite{hunter_fullPaper} &   static analysis&integrated	&  \cmark{} \\
        CodeCarbonCopy \cite{code_carbon_copy} &   dynamic analysis&external	&  \cmark{} \\
        \citet{smartPaste} &   static analysis&external	&  \cmark{} \\
        \citet{a_study_on_tree-based_code_extraction} &   static analysis&not mentioned	&  \cmark{} \\
        \citet{an_initial_step_toward_organ_transplantation} &   -&-	&  -  \\
        \citet{genetic_improvement_survey} &   -&-	&  -  \\
        Splicer \cite{program_splicing} &   static analysis&integrated	&  \cmark{} \\
        \citet{justyna_specialise_downstream} &   dynamic analysis&integrated	&  \cmark{} \\
        \citet{a_study_on_code_transplantation_technique} &   static analysis&not mentioned	&  \cmark{} \\
        AppTestMigrator \cite{automated_test_migration_for_mobile_apps}&   dynamic analysis&external	&  \cmark{} \\
        GUITestMigrator \cite{test_migration_for_efficient_large-scale_assessment_of_mobile_app_coding_assignments}&   dynamic analysis&external	&  \cmark{} \\
        \citet{dissection_on_java_organs_in_github_repositories} &   -&-	&  -\\
        CraftDroid \cite{test_transfer_across_mobile_apps_through_semantic_mapping}&   dynamic analysis&external	&  \cmark{} \\
        TestMig \cite{testmig_migrating_gui_test_cases_from_ios_to_android}&   dynamic analysis&external	&  \cmark{} \\
        AppTestMigrator \cite{test_migration_between_mobile_apps_with_similar_functionality}&   dynamic analysis&external	&  \cmark{} \\
        NLP2TestableCode \cite{NLP2TestableCode} &   static analysis&integrated	&  \cmark{} \\
        \citet{semi-automated_test-case_propagation_in_fork_ecosystem} &  -&-  	&  - \\
        \citet{test_transplantation_through_dynamic_test_slicing} &  dynamic analysis &not mentioned 	&  \cmark{} \\
        AdaptivePaste \cite{adaptivePaste} &  static analysis &external 	& \cmark{} \\
        \ts{TransDroid \cite{gui_test_transfer_from_web_to_android}} & \ts{dynamic analysis} & \ts{external} & \ts{\cmark{}}\\
        \citet{software_testing_research_challenges_an_industrial_perspective} &  - &- 	&  - \\
        ProdScalpel \cite{prodScalpel} &  dynamic analysis&integrated 	&  \cmark{} \\
        \citet{syntax_preserving_program_slicing} &   static analysis&external	&  \cmark{}\\
        \ts{LLMigrate \cite{automated_test_transfer_across_android_apps_using_large_language_models}} & \ts{dynamic analysis} & \ts{external} & \ts{\cmark{}}\\
        \hline
    \end{tabular}
\end{table*}

\begin{table*}
    \caption{Publications on automated software transplantation, with summary of answers to our research questions (RQ6-RQ8). Tool names added, if provided in a given publication.}
    \label{tab:papers_summary_rq6_to_rq8}
    \centering
    \begin{tabular}{llll}
        \hline
        Tool Name / Paper & RQ6: organ type & RQ7: user study &RQ8: tool\\
        \hline
        Code Conjurer \cite{code_conjurer}& feature 	& - &\cmark{}\\
        Jigsaw \cite{jigsaw_fullPaper} & feature 	& \cmark{} &\cmark{}\\
        Jigsaw \cite{jigsaw_2pages} & feature 	& - &\cmark{}\\
        \citet{semi_automating_pragmatic_reuse_tasks} & feature 	& - &\cmark{}\\
        Procrustes \cite{procrustes} & feature 	&\cmark{}  &\cmark{}\\
        Gilligan \cite{pragmatic_software_reuse_PhD} & feature 	& \cmark{} &\cmark{}\\
        \citet{reusing_web_application_user-interface_controls} & feature 	& - &\cmark{}\\
        Gilligan \cite{gilligan} & feature 	&\cmark{} &\cmark{}\\
        \citet{genetic_programming_for_reverse_engineering} & - 	& - &-\\
        \citet{towards_automatic_client-side_feature_reuse} & feature 	& - &\cmark{}\\
        \citet{SBSE_for_SPL} & - 	& - &-\\
        \citet{justyna_specialise_minisat} & feature 	& - & \cmark{}\\
        Grow and Graft \cite{babel_pidgin} & feature 	& - &\cmark{}\\
        \(\mu Scalpel\) \cite{muScalpel} & feature 	& - &\cmark{}\\
        \citet{towards_specifying_pragmatic_software_reuse} & feature 	& - &-\\
        \(\mu Scalpel\) \cite{muScalpel_kate_ide} & feature 	& - &\cmark{}\\
        Grow and Serve \cite{grow_django_citation} & feature 	& - &\cmark{}\\
        \citet{program_fracture_and_recombination_for_efficient_automatic_code_reuse} & feature 	& - &\cmark{}\\
        \citet{horizontal_code_transfer_via_program_fracture_and_recombination} & feature 	& - &\cmark{}\\
        Hunter \cite{hunter} & feature 	& \cmark{} &\cmark{}\\
        Hunter \cite{hunter_fullPaper} & feature 	&\cmark{}  &\cmark{}\\
        CodeCarbonCopy \cite{code_carbon_copy} & feature 	& - &\cmark{}\\
        \citet{smartPaste} & feature 	& - &\cmark{}\\
        \citet{a_study_on_tree-based_code_extraction} & feature 	& - &\cmark{}\\
        \citet{an_initial_step_toward_organ_transplantation} & feature 	& - &-\\
        \citet{genetic_improvement_survey} & - 	& - &-\\
        Splicer \cite{program_splicing} & feature 	& \cmark{} &\cmark{}\\
        \citet{justyna_specialise_downstream} & feature & - & \cmark{}\\
        \citet{a_study_on_code_transplantation_technique} & feature 	& - &\cmark{}\\
        AppTestMigrator \cite{automated_test_migration_for_mobile_apps}& test 	& - &\cmark{}\\
        GUITestMigrator \cite{test_migration_for_efficient_large-scale_assessment_of_mobile_app_coding_assignments}& test 	& - &\cmark{}\\
        \citet{dissection_on_java_organs_in_github_repositories} & feature 	& - &-\\
        CraftDroid \cite{test_transfer_across_mobile_apps_through_semantic_mapping}& test 	& - &\cmark{}\\
        TestMig \cite{testmig_migrating_gui_test_cases_from_ios_to_android}& test 	& - &\cmark{}\\
        AppTestMigrator \cite{test_migration_between_mobile_apps_with_similar_functionality}& test 	& - &\cmark{}\\
        NLP2TestableCode \cite{NLP2TestableCode} & feature 	& - &\cmark{}\\
        \citet{semi-automated_test-case_propagation_in_fork_ecosystem} & test 	& - &-\\
        \citet{test_transplantation_through_dynamic_test_slicing} & test 	& - &\cmark{}\\
        AdaptivePaste \cite{adaptivePaste} & feature 	& \cmark{} &\cmark{}\\
        \ts{TransDroid \cite{gui_test_transfer_from_web_to_android}} & \ts{test} & \ts{-} & \ts{\cmark{}}\\
        \citet{software_testing_research_challenges_an_industrial_perspective} & - 	& - &-\\
        ProdScalpel \cite{prodScalpel} & feature 	& \cmark{}  &\cmark{}\\
        \citet{syntax_preserving_program_slicing} & feature 	& - &\cmark{}\\
        \ts{LLMigrate \cite{automated_test_transfer_across_android_apps_using_large_language_models}} & \ts{test} & \ts{-} & \ts{\cmark{}}\\
        \hline
    \end{tabular}
\end{table*}

\subsection{RQ1: Locating An Organ} \label{sec:rq_organ_location}
\begin{figure}
    \centering
    \includegraphics[width=0.6\linewidth]{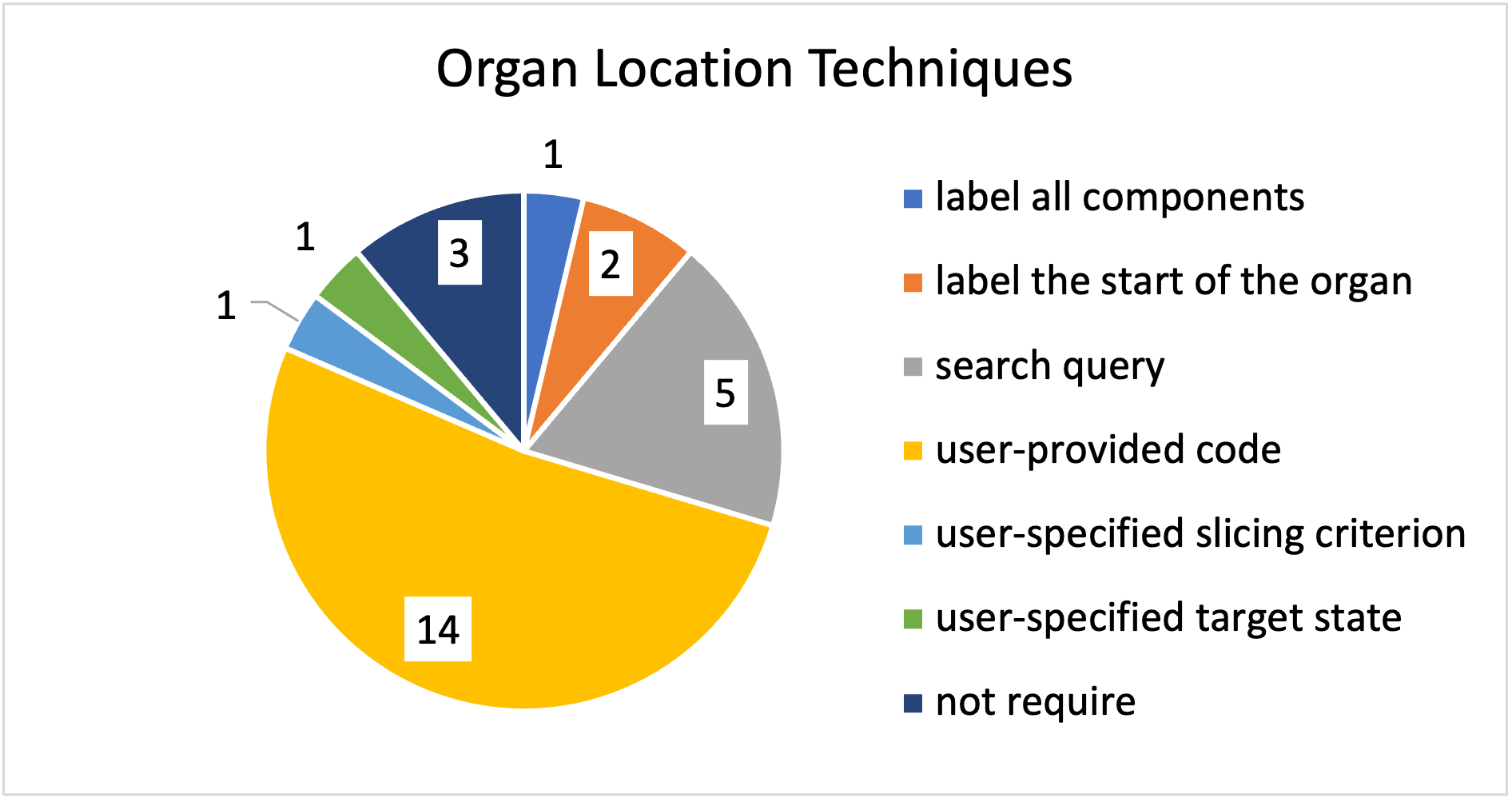}
    \caption{Organ location strategies in \totalNumberOfDistinctTransplantationApproaches{} distinct automated software transplantation approaches: \numberOfApproachesLabelAllReuseDecision{} strategy requires the user to manually label all software components with a reuse decision; \numberOfApproachesUserProvidesSearchQuery{} approaches require the user to provide a search query; \numberOfApproachesUserProvidesOrganLocation{} approaches require lightweight information, such as an organ's entry point; the remaining \numberOfApproachesNotRequireOrganLocation{} approaches do not require upfront information on organ location.}
    \Description{}
    \label{fig:organ_location_technique}
\end{figure}

Once the functionality of interest, i.e., an organ, has been decided, a transplantation approach must locate it in the donor software before its migration into the target, i.e., host, application.
From our literature review, we found several techniques proposed for this task. 
We present techniques for organ location from the least to the most automated ones. 
Our literature review revealed \totalNumberOfDistinctTransplantationApproaches{} distinct approaches.
Figure \ref{fig:organ_location_technique} shows the distribution of the different ways each software transplantation approach locates an organ.

Four publications rely on the \textit{pragmatic reuse plan} to locate an organ.
Creating the plan is a manual process, requiring the user to explicitly label each software component with the reuse decision. 
For example, the Procrustes~\cite{procrustes} and Gilligan~\cite{gilligan} approaches both use a reuse plan to identify software components for transplantation.

Another way of locating an organ is for users to manually provide limited information about its location. In this setting, \(\mu Scalpel\)~\cite{muScalpel, muScalpel_kate_ide} and ProdScalpel~\cite{prodScalpel} require users to explicitly label the entry point of an organ's functionality using a special comment.

A more direct form of manual specification involves users explicitly providing the organ's code. This includes approaches from the copy-and-paste domain, such as AdaptivePaste~\cite{adaptivePaste}, Jigsaw~\cite{jigsaw_2pages, jigsaw_fullPaper}, and \citet{smartPaste}, where users manually select a target code snippet from a donor application and paste it into the desired location in the host application. Similarly, CodeCarbonCopy~\cite{code_carbon_copy} and the transplantation techniques proposed by \citet{a_study_on_tree-based_code_extraction, a_study_on_code_transplantation_technique} rely on the user identifying and supplying the donor code fragment that will be transplanted. 

This category also includes test case transplantation approaches, which rely on user-supplied test artifacts. TestMig~\cite{testmig_migrating_gui_test_cases_from_ios_to_android} requires the user to provide an iOS test suite, which is then transplanted to an Android application that implements the same functionality as the original iOS app. \ts{TransDroid~\cite{gui_test_transfer_from_web_to_android} requires a web application's test suite, which is transplanted to an Android application implementing the same functionality.} AppTestMigrator~\cite{test_migration_between_mobile_apps_with_similar_functionality, automated_test_migration_for_mobile_apps}, CraftDroid~\cite{test_transfer_across_mobile_apps_through_semantic_mapping}, GUITestMigrator~\cite{test_migration_for_efficient_large-scale_assessment_of_mobile_app_coding_assignments}, \ts{and LLMigrate~\cite{automated_test_transfer_across_android_apps_using_large_language_models}} require Android test suites as input and transplant them into other Android applications with similar functionality. 

Similarly,~\citet{syntax_preserving_program_slicing} proposed syntax-preserving program slicing for software transplantation, where the user is required to supply one of the following details for an organ:
\begin{enumerate}
    \item Location: specific statement or line of code;
    \item Configuration option: if code only for a specific configuration is required;
    \item Identifier: an identifier that encapsulates the function of interest;
    \item String: any string that describes the function of interest.
\end{enumerate}
The chosen slicing criterion is later used to create a program slice that implements the feature of interest.

Beyond supplying code or slicing criteria, some approaches require users to specify higher-level information about the organ. For example, \citet{test_transplantation_through_dynamic_test_slicing}'s approach requires users to specify a target state, which is then used to backtrace program execution and extract the corresponding organ. In the context of web applications, \citet{reusing_web_application_user-interface_controls, towards_automatic_client-side_feature_reuse, DBLP:phd/se/Maras14} require developers to specify a UI control in the donor application.

Other software transplantation approaches locate organs by relying on a code search engine as part of the transplantation process. In these approaches, organ location is performed by issuing a query over a database of code snippets and selecting candidate components from the returned results. However, \citet{code_conjurer} found that early code search techniques (e.g., keyword, signature, and name-based matching) are often imprecise and tend to return irrelevant results.

To address these limitations, Code Conjurer~\cite{code_conjurer} introduced a software transplantation approach that uses a code search engine in which queries are expressed in the style of a programming language interface, enabling the engine to search for source code with higher precision. Similarly, Splicer~\cite{program_splicing} is a transplantation approach that constructs code search queries from a draft program and associated natural language comments to locate candidate organs from its program database, which is constructed from the SOURCERER~\cite{sourcer} source code repository. NLP2TestableCode~\cite{NLP2TestableCode} is another transplantation approach that uses a code search engine to locate organs by querying an offline Stack Overflow database based on user-provided natural language descriptions. Hunter~\cite{hunter, hunter_fullPaper} likewise employs code search for organ location, requiring users to provide a method signature, an English description, and test cases to retrieve relevant code snippets.

\citet{program_fracture_and_recombination_for_efficient_automatic_code_reuse, horizontal_code_transfer_via_program_fracture_and_recombination} introduced an approach known as \textit{program fracture and recombination}. This approach fractures one or more donor programs into small fragments (i.e., shards) at varying levels of granularity to construct a database of organs. To locate a replacement shard for a host program, the original shard is executed within the host to record its input–output behaviour. Each shard in the organ database is then executed in its original program context, and those exhibiting semantically compatible behaviour are identified as potential matches. However, the authors do not describe how organs are located for shard insertion, where the goal is to introduce new functionality that is not already present in the host program.

In addition, \numberOfApproachesNotRequireOrganLocation{} approaches do not require the location of an organ as the first step. 
For instance,~\citet{justyna_specialise_downstream, justyna_specialise_minisat} focused on software specialisation. 
Here, it is not known upfront which organs might be useful to specialise to the downstream task. 
The approach relies on a `code bank', which is prepared by the authors. 
Code bank includes all source code from the host program itself as well as the code that is present in other versions of the code.
In the \textit{grow and graft/serve} approaches~\cite{grow_django_citation, babel_pidgin}, the organ is grown from scratch with the guidance of user inputs, such as a list of API names and pre-defined fitness functions.

\begin{mdframed}
\textbf{Answer to RQ1: Most software transplantation techniques require varying levels of assistance from software engineers when locating an organ.
Most manual approaches involve engineers explicitly labelling each component with a reuse decision (\percentageOfApproachesLabelAllReuseDecision{}), while automated methods only require engineers to provide lightweight information about the organ's location (\percentageOfApproachesUserProvidesOrganLocation{}) or provide a search query for a code search engine (\percentageOfApproachesUserProvidesSearchQuery{}).
Lastly, \percentageOfApproachesNotRequireOrganLocation{} of approaches do not require upfront information on the organ's location.
}
\end{mdframed}

\subsection{RQ2: Organ Extraction} \label{sec:rq3_organ_extraction}
\begin{figure}
    \centering
    \includegraphics[width=0.6\linewidth]{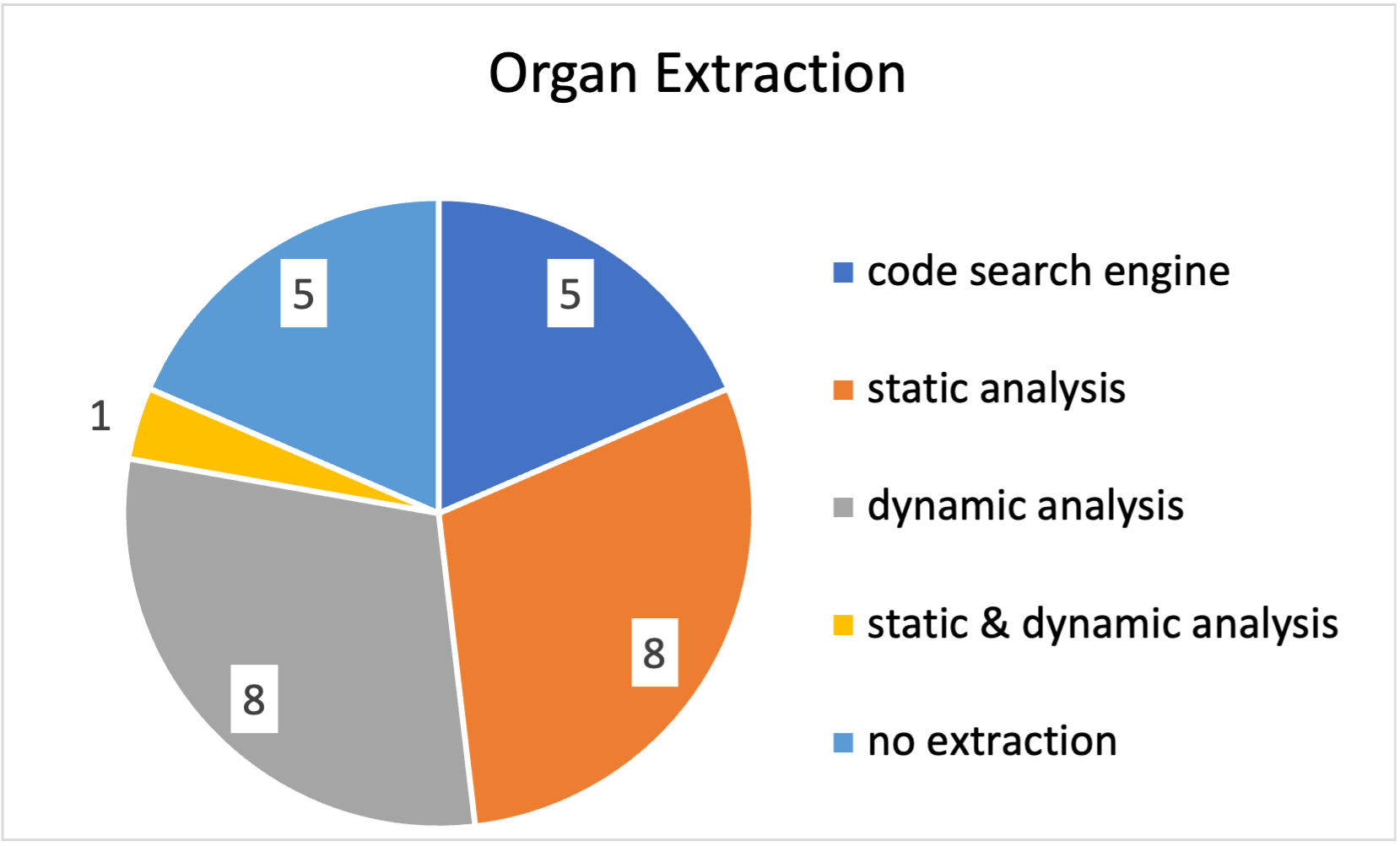}
    \caption{\totalNumberOfDistinctTransplantationApproaches{} distinct transplantation approaches were proposed. \percentageOfApproachesUsingCodeSearchEngineForOrganExtraction{} of approaches use a code search engine to extract an organ, while \percentageOfApproachesUsingStaticAnalysisForOrganExtraction{} and \percentageOfApproachesUsingDynamicAnalysisForOrganExtraction{} of them use static and dynamic analysis techniques, respectively. \percentageOfApproachesUsingStaticAndDynamicAnalysisForOrganExtraction{} of approaches utilise both static and dynamic analysis within a single approach, such as \citet{reusing_web_application_user-interface_controls}.
    \percentageOfApproachesNoOrganExtraction{} of the approaches do not have an organ extraction phase, as the user is required to manually provide the organ \ts{\notePercentageRounding{}.}}
    \Description{}
    \label{fig:organ_extraction_technique}
\end{figure}

After identifying the location of an organ in the donor, the next step is to extract it. 
This section discusses the organ extraction techniques used by the software transplantation approaches.
We categorise these techniques into three groups: code search, static analysis, and dynamic analysis.
Figure \ref{fig:organ_extraction_technique} shows the distribution of the different ways each software transplantation approach extracts an organ.

\begin{table}[h]
    \centering
    \begin{tabular}{cp{6.5cm}}
    \hline
    Section & Approach Name / Tool Name \\
    \hline
    \ref{sec:rq_organ_extraction_dynamic_analysis} & AppTestMigrator~\cite{test_migration_between_mobile_apps_with_similar_functionality,automated_test_migration_for_mobile_apps} \\
    \ref{sec:rq_organ_extraction_static_analysis} & CodeCarbonCopy~\cite{code_carbon_copy} \\
    \ref{sec:rq3_code_search} & Code Conjurer~\cite{code_conjurer} \\
    \ref{sec:rq_organ_extraction_dynamic_analysis} & CraftDroid~\cite{test_transfer_across_mobile_apps_through_semantic_mapping} \\
    \ref{sec:rq_organ_extraction_static_analysis} & Gilligan / Procrustes~\cite{gilligan,procrustes,pragmatic_software_reuse_PhD,semi_automating_pragmatic_reuse_tasks} \\
    \ref{sec:rq_organ_extraction_dynamic_analysis} & GUITestMigrator~\cite{test_migration_for_efficient_large-scale_assessment_of_mobile_app_coding_assignments} \\
    \ref{sec:rq3_code_search} & Hunter~\cite{hunter, hunter_fullPaper} \\
    \ts{\ref{sec:rq_organ_extraction_dynamic_analysis}} & \ts{LLMigrate~\cite{automated_test_transfer_across_android_apps_using_large_language_models}} \\
    \ref{sec:rq3_code_search} & NLP2TestableCode~\cite{NLP2TestableCode} \\
    \ref{sec:rq_organ_extraction_static_analysis} & ProdScalpel~\cite{prodScalpel} \\
    \ref{sec:rq3_code_search} & Program Fracture and Recombination~\cite{program_fracture_and_recombination_for_efficient_automatic_code_reuse, horizontal_code_transfer_via_program_fracture_and_recombination} \\
    \ref{sec:rq_organ_extraction_dynamic_analysis} & SMALL-MINCE~\cite{test_transplantation_through_dynamic_test_slicing} \\
    \ref{sec:rq3_code_search} & Splicer~\cite{program_splicing} \\
    \ref{sec:rq_organ_extraction_dynamic_analysis} & TestMig~\cite{testmig_migrating_gui_test_cases_from_ios_to_android} \\
    \ts{\ref{sec:rq_organ_extraction_dynamic_analysis}} & \ts{TransDroid~\cite{gui_test_transfer_from_web_to_android}} \\
    \ref{sec:rq_organ_extraction_static_analysis} & \(\mu Scalpel\)~\cite{muScalpel,muScalpel_kate_ide} \\
    \ref{sec:rq_organ_extraction_static_analysis} & \citet{syntax_preserving_program_slicing} \\
    \ref{sec:rq_organ_extraction_static_analysis} & \citet{a_study_on_code_transplantation_technique} \\
    \ref{sec:rq_organ_extraction_static_analysis} & \citet{a_study_on_tree-based_code_extraction} \\
    \ref{sec:rq_organ_extraction_dynamic_analysis} & \citet{reusing_web_application_user-interface_controls} \\
    \ref{sec:rq_organ_extraction_dynamic_analysis} & \citet{towards_automatic_client-side_feature_reuse, towards_specifying_pragmatic_software_reuse, DBLP:phd/se/Maras14} \\
    \ref{sec:rq_organ_extraction_static_analysis} & \citet{justyna_specialise_downstream,justyna_specialise_minisat} \\    
    \hline
    \end{tabular}
    \caption{Organ extraction techniques employed by \totalNumberOfDistinctTransplantationApproaches{} distinct software transplantation approaches. We only show those that have an organ extraction phase.}
    \label{tab:distinct_software_transplantation_approaches_organ_extraction}
\end{table}

\subsubsection{Code Search} \label{sec:rq3_code_search}
Several proposed automated software transplantation approaches use a code search engine to extract an organ.
A code search engine stores and returns code snippets based on user queries from selected program databses.
Consequently, the organ extraction phase for approaches that use code search is simply the process of sending a query to the program database and receiving a set of code snippets in return.

Code Conjurer~\cite{code_conjurer} is one of the first transplantation tools to use code search. 
It forms a query based on the user-provided test suite by extracting the interface from the test suite and using it to generate a search query. 
Often, queries involving complex interfaces return little to no results from code searches. 
To address this, \citet{code_conjurer} developed the \textit{Automated Adaptation Engine}, which returns components as long as they provide the required functionality. 
This works by permuting all possible mappings between the test suite's and the candidate's interface.

Interface-based queries are not always effective, as a component's interface may not provide enough information to accurately describe the desired organ. 
This can lead to incorrect or no results being returned. 
A more effective alternative is to use natural language descriptions of the desired organ. 
For instance, transplantation tools such as Hunter~\cite{hunter, hunter_fullPaper} and NLP2TestableCode~\cite{NLP2TestableCode} use user-provided English descriptions to formulate queries for code search engines. 
Similarly, Splicer~\cite{program_splicing} combines natural language input with features extracted from a draft program to construct a query.

Although natural language descriptions can improve the retrieval of relevant code snippets, the burden remains on the user to clearly and accurately describe the desired organ, which introduces a risk of ambiguity and misinterpretation. An alternative approach is to locate candidate organs based on behavioural specifications, such as input-output pairs. Program Fracture and Recombination~\cite{horizontal_code_transfer_via_program_fracture_and_recombination, program_fracture_and_recombination_for_efficient_automatic_code_reuse} adopts this strategy by executing a reference shard to capture its input-output behaviour and using this information to search a database of candidate shards. Each candidate is executed and compared against the behavioural specification. Those shards that exhibit semantically compatible behaviour are returned as potential organs. Compared to interface- or natural-language-based queries, behavioural matching can reduce ambiguity by grounding the search process in observable program semantics.

Nonetheless, code search approaches are limited by the availability of the desired organ in the search engine's program database. 
If the organ is not present, these methods fail to extract it. 
A more reliable alternative is to use program analysis techniques, static or dynamic, where the location of the organ is known and can be extracted directly.

\subsubsection{Static Analysis} \label{sec:rq_organ_extraction_static_analysis}
Static analysis refers to the examination of the source code without executing it, allowing for the identification and extraction of specific code segments (i.e., organs) based on predefined criteria or patterns. 
Static analysis techniques can provide deeper insights into the code structure, dependencies, and relationships between different components. 
While code search engines are useful for locating candidate components quickly, static analysis offers greater precision in understanding how those components are constructed and interconnected.

Early transplantation approaches that used static analysis for code extraction, such as Gilligan~\cite{gilligan} and Procrustes~\cite{procrustes, pragmatic_software_reuse_PhD}, relied on manually specified pragmatic reuse plans to move selected components into the host program based on reuse decisions. 
Given the reuse plan, an approach automatically extracts the corresponding code fragments from the donor program into the host. 
Both approaches maintain the original package hierarchy for each of comprehension. 
Although the extraction is automatic, there is a risk of error as the extraction is guided by human judgement during the creation of the pragmatic reuse file.

A better alternative is to use program slicing techniques, which automatically extract relevant code segments based on user-defined criteria, such as an approach from \citet{a_study_on_code_transplantation_technique} it uses JavaSlicer to slice an organ based on the user-defined organ location.

\citet{syntax_preserving_program_slicing} proposed a slicing-based organ extraction technique that preserves the original program syntax, including C preprocessor directives such as \#ifdef. 
The method is built on a specialised code representation called the \textit{variability-aware Code Property Graph} (vaCPG), which captures both structural and lexical elements of the code. Organ extraction is performed by slicing the vaCPG using user-defined criteria, ensuring that the resulting code fragment maintains syntactic integrity.

\begin{figure}
    \centering
    \includegraphics[width=0.6\linewidth]{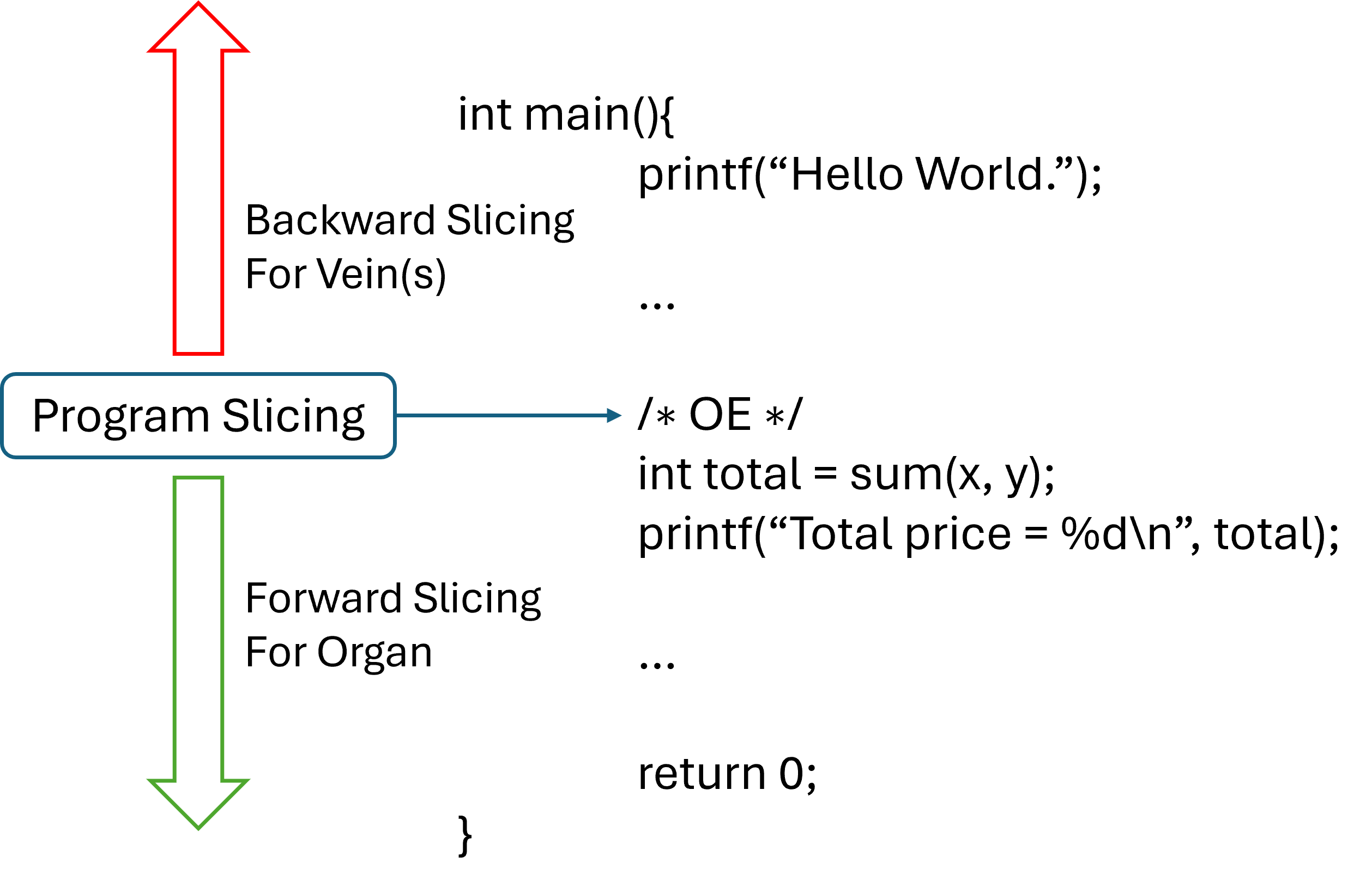}
    \caption{An example organ extraction step of \(\mu Scalpel\) \cite{muScalpel}. The approach requires the user to label the start of an organ with a special comment, where it later applies the program slicing technique to extract an organ and a vein via forward and backward slicing, respectively.}
    \Description{}
    \label{fig:muScalpel_organ_extraction}
\end{figure}

Similarly, \(\mu Scalpel\) \cite{muScalpel,muScalpel_kate_ide} uses forward and backward slicing to extract an organ from a donor program (see Figure \ref{fig:muScalpel_organ_extraction}). 
Given a user-annotated organ entry point, it performs forward slicing to produce an over-approximation of the organ—referred to as the \textit{over-organ}—which includes all necessary helper code to make the organ self-contained. 
It then applies backward slicing from the main function to identify a path that constructs the required program environment for the organ's execution. 
This path is known as a \textit{vein}. 
Theoretically, multiple veins may exist, and any vein can be selected as long as it correctly builds the programming environment.

In their subsequent work, \citet{prodScalpel} also use the same approach (e.g., forward and backward slicing) to extract an organ.

In contrast to slicing-based extraction techniques, CodeCarbonCopy~\cite{code_carbon_copy} performs organ extraction using dependency-based static analysis. 
Starting from a developer-specified function, the approach statically identifies all code elements required for its compilation and execution. 
To this end, CodeCarbonCopy constructs a compile-time dependence graph by tracing both compile-time dependencies and statically identifiable run-time dependencies, where nodes represent code elements and edges represent compile-time dependencies. 
All reachable nodes in the graph are collected and emitted into a single source file, treating the organ as a callable function from the host program. 
During extraction, accesses to global variables in the donor environment are identified and prepared for lifting, enabling their replacement with explicit function parameters during the subsequent adaptation phase.

Lastly, some transplantation approaches rely on static analysis for organ extraction. 
In the multi-donor transplantation approach proposed by \citet{justyna_specialise_downstream, justyna_specialise_minisat}, organs are extracted by automatically identifying syntactic differences between donor variants and the host program. 
These differences are explicitly isolated as conditional code fragments and extracted from the donor code. 
The extracted code fragment and original host code form a shared repository called `code bank', which serves as a pool of code elements available for subsequent transplantation steps.

\ts{\citet{a_study_on_tree-based_code_extraction}} present a manual approch for organ extraction.  
The authors manually analyse the AST of the donor program to identify the organ and its dependencies.

Moreover, we note here that one approach relies on both static and dynamic analysis techniques to extract an organ, namely \citet{reusing_web_application_user-interface_controls}.
We further discuss this approach in the next section.

\subsubsection{Dynamic Analysis} \label{sec:rq_organ_extraction_dynamic_analysis}
Another approach to organ extraction is the use of dynamic analysis techniques.
Dynamic analysis involves executing the program and monitoring its behaviour at runtime to identify and extract specific code segments (i.e., organs).
This enables the extraction of only the code that is actually executed in response to specific scenarios or inputs, increasing precision and relevance of the extracted code.
In contrast, static analysis often produces an over-approximation, including code that may never be executed.
Similarly, a code search engine typically focuses on syntactic matching and lacks the semantic context that dynamic analysis can provide.

Early transplantation work that uses dynamic analysis for organ extraction was semi-automated. 
For example, \citet{reusing_web_application_user-interface_controls} extract a user-interface control (organ) from a web application using a combination of manual specification and automated analysis. First, user manually identifies the target HTML node corresponding to the organ. Then, user interacts with the organ while the approach records and extracts all relevant code and resources based on these interactions. Based on this execution trace, the approach applies static analysis to remove unused HTML, CSS, and JavaScript code, producing a self-contained organ. However, this approach requires the user to manually specify the corresponding HTML node (organ's location) in the web application, which can be time-consuming and error-prone.

In their subsequent work~\cite{towards_automatic_client-side_feature_reuse, towards_specifying_pragmatic_software_reuse}, the authors utilise a client-side dependency graph instead of interaction recording.

Test case transplantation approaches, on the other hand, execute donor test cases to generate execution traces or capture GUI events and assertions, which are then used to identify relevant code. 
For example, SMALL-MINCE~\cite{test_transplantation_through_dynamic_test_slicing} builds an acyclic graph from execution traces and traverses it backward from the target node to extract the organ. 
Similarly, mobile-specific approaches such as GUITestMigrator~\cite{test_migration_for_efficient_large-scale_assessment_of_mobile_app_coding_assignments}, AppTestMigrator~\cite{test_migration_between_mobile_apps_with_similar_functionality}, and CRAFTDROID~\cite{test_transfer_across_mobile_apps_through_semantic_mapping} extract GUI-related organs on Android. \ts{Similarly, TestMig~\cite{testmig_migrating_gui_test_cases_from_ios_to_android} and TransDroid~\cite{gui_test_transfer_from_web_to_android} migrate GUI test cases from iOS and web applications, respectively, to Android applications. Unlike prior event-based approaches, LLMigrate~\cite{automated_test_transfer_across_android_apps_using_large_language_models} uses an LLM to abstract executed donor tests into natural language representations of their functional intent.}

\begin{mdframed}
\textbf{
Answer to RQ2: \percentageOfApproachesUsingStaticAnalysisForOrganExtraction{} and \percentageOfApproachesUsingDynamicAnalysisForOrganExtraction{} of software transplantation approaches use static and dynamic analysis for extracting an organ, respectively.
A further \percentageOfApproachesUsingStaticAndDynamicAnalysisForOrganExtraction{} of approaches combine both static and dynamic analysis within a single extraction strategy.
In addition, \percentageOfApproachesUsingCodeSearchEngineForOrganExtraction{} of approaches use code search instead.
Lastly, \percentageOfApproachesNoOrganExtraction{} of software transplantation approaches have no explicit organ extraction phase, as the user is required to manually provide an organ.
}
\end{mdframed}

\subsection{RQ3: Organ Adaptation for Implantation} \label{sec:rq_organ_adaptation}

The goal of the organ adaptation phase is to tailor the extracted organ into the host environment at the implantation location. 
This involves identifying and resolving dependencies between the organ and the host. 
In this section, we present the methods employed by the software transplantation approaches for the organ adaptation task by grouping similar approaches together. 
We divided these into two types: static and dynamic.

\subsubsection{Static Analysis} \label{sec:rq_organ_adaptation_static_analysis}
In the context of organ adaptation, static analysis is used to identify organ's dependencies and resolve them in the host's context.

Early work on static analysis for organ adaptation was manual~\citet{a_study_on_code_transplantation_technique, a_study_on_tree-based_code_extraction}.
Later approaches introduced automated support for organ adaptation.
Yet, it primarily focuses on resolving trivial compilation errors, such as missing import statements or undefined variables.
For instance, \citet{gilligan, pragmatic_software_reuse_PhD, procrustes} leverage the pragmatic reuse plan—described in Section~\ref{sec:rq_organ_extraction_static_analysis}—to guide adaptation. 
Similarly, Code Conjurer~\cite{code_conjurer} works by resolving dependencies (e.g., adding missing import statements), while NLP2TestableCode~\cite{NLP2TestableCode} applies heuristic rules to relocate imports and fix syntactic issues. 
Nonetheless, these methods only focus on syntactic correctness and do not guarantee semantic correctness of the transplanted organ.

\citet{hunter} argued that resolving dependencies alone is insufficient to ensure that an organ functions correctly in a new context, as mismatches may exist between the interface required by the host program and that provided by the organ.
With this in mind, \citet{hunter, hunter_fullPaper} proposed Hunter, which uses integer linear programming (ILP) to compute an optimal type-based alignment between the host interface and candidate organs.
However, Hunter can only succeed when a type-compatible alignment exists between the host interface and a candidate organ that enables the generation of adapter code to correctly invoke the organ.
When no such alignment exists, the adaptation fails.

On the other hand, Splicer~\cite{program_splicing} adapts an organ by enumerating codelets (e.g., expressions and statements) extracted from code snippets returned by a code search engine and inserting them into holes in a draft program. 
Candidate codelets are selected based on type compatibility with the surrounding context and are further adapted through reference substitution, where variable names are renamed to resolve undefined references in the host program. 
Similar to Hunter, Splicer may fail to produce a valid transplant if none of the retrieved code snippets can be adapted to satisfy the host program's typing, reference-resolution, and correctness constraints. 
Moreover, given that Splicer enumerates through all returned code snippets, its runtime can be substantial, hindering its adoption in practice.

Other transplantation approaches have explored variable renaming at the organ adaptation stage. 
Both \citet{smartPaste}'s approach and AdaptivePaste~\cite{adaptivePaste} leverage machine learning models to rename variables in a way that aligns with the host program's naming conventions and semantics.

Moreover, there are two variants of AdaptivePaste: Uni-decoder Transformer (i.e., AP-uni) and Parallel-decoder transformer (i.e., AP-parallel). 
AP-uni adapts all identifiers in a pasted snippet together, allowing each prediction to depend on previously adapted names. This joint adaptation generally results in higher overall accuracy. In contrast, AP-parallel adapts each identifier independently, which enables per-variable confidence estimation and allows uncertain identifiers to be left unadapted.
As a result, neither variant strictly outperforms the other: AP-uni achieves higher overall accuracy, while AP-parallel is more robust when some identifiers are difficult to adapt or when partial adaptation is preferred.

However, the aforementioned variable renaming strategies have some limitations.
For instance, \citet{smartPaste}'s approach cannot rename variables that are not explicitly initialised, while AdaptivePaste struggles when a variable has been reassigned a new name. 
As with any machine learning-based approach, there is an inherent risk that these models may fail to generalise to unseen code snippets.

Moreover, other transplantation approaches adapt an organ to the host context by adding only the missing lines of code and files from the donor, based on a comparison between the organ and the implantation site. 
As a result, these methods are limited to transplanting between software systems that share a certain level of similarity. 
For example, Jigsaw~\cite{jigsaw_fullPaper, jigsaw_2pages} is an Eclipse IDE plug-in where users select a method 
from the donor program as the organ and specify a target method or class in the host as the implantation site. 
The tool analyses an augmented abstract syntax tree called the Correspondence AST (CAST) for both the organ and the host, and copies only those nodes from the organ that have no corresponding match in the host. 
Yet, it requires user intervention to resolve conflicts, such as when a single node from the organ maps to multiple nodes in the host.

In contrast, the approach by \citet{syntax_preserving_program_slicing} operates at the repository level rather than within an IDE, allowing the addition of files and code lines across closely related software projects, such as those found in forked repositories. 
During organ adaptation, both the extracted slice and the corresponding portion of the host are transformed into a common text-based markup representation and compared using exact syntactic equality. 
That is, a code fragment is considered present in the host only if its markup representation matches the slice exactly, without applying any heuristics or similarity measures. 
As a result, only files and lines that are strictly identical are matched, and all unmatched fragments are treated as missing and copied into the host.

Beyond adapting functionality in traditional programming languages like Python, Java, and C, some studies explored reuse in the context of web technologies, including HTML, CSS, and JavaScript. For instance, \citet{reusing_web_application_user-interface_controls} introduced a method for reusing web UI components by addressing JavaScript library conflicts and naming clashes. When automatic resolution was insufficient, the system required manual user input. However, being based solely on static analysis, the approach struggled to detect conflicts that emerge only at runtime.

To address this limitation, in their follow-up work, \citet{towards_automatic_client-side_feature_reuse} proposed a more advanced method that eliminates the need for user intervention. This approach incorporates dynamic analysis outputs—such as dependency graphs and execution traces—to better detect and manage runtime conflicts. Still, a major limitation persists: the tool cannot fully resolve certain dynamic conflicts, particularly complex cases involving JavaScript's built-in prototype extensions, as illustrated in one of their case studies \cite{towards_automatic_client-side_feature_reuse}. 

Although static analysis has served as a foundational technique for organ adaptation, it has notable limitations in precision and in handling complex dependencies. While incorporating dynamic analysis outputs into static analysis can improve conflict resolution, it still falls short in addressing complex runtime scenarios. To overcome these limitations, dynamic analysis could be employed for organ adaptation.

\subsubsection{Dynamic Analysis} \label{sec:rq_organ_adaptation_dynamic_analysis}
To dynamically adapt an organ, an approach must execute the organ within the host's environment.
This ensures that the organ is successfully integrated and that the host can exercise the desired functionalities provided by the organ.
However, this approach inherits the inherent limitation of dynamic analysis: it only verifies the features that are actually executed during runtime.
Consequently, it provides an under-approximation of correctness, as unexecuted code paths remain unchecked.

Early work in this area used Genetic Programming (GP) to independently grow an organ from scratch before moving it into the host program. \citet{babel_pidgin} introduced the \textit{grow and graft} approach, which incorporates user-provided guidance (e.g., API names, fitness functions, constraint ordering, and other relevant inputs) to guide the GP. After the organ is grown, an approach enumerates through possible insertion points within the host, evaluating each for compilation, crash, and test case correctness.

In their subsequent work, \textit{grow and graft}~\cite{grow_django_citation}, the organ is similarly grown from scratch. However, instead of being injected into the host, the organ is used as an external module that interacts with the host system.

Later on, \citet{justyna_specialise_downstream, justyna_specialise_minisat} introduced multi-donor software transplantation, where an organ can originate from both the donor and the host to support software specialisation. The authors use GP to perform adaptation tasks such as deletion, replacement, and copying of code fragments. In their approach, the donor and host are different versions of the same program, enabling more targeted and efficient specialisation.

\begin{figure}
    \centering
    \includegraphics[width=0.6\linewidth]{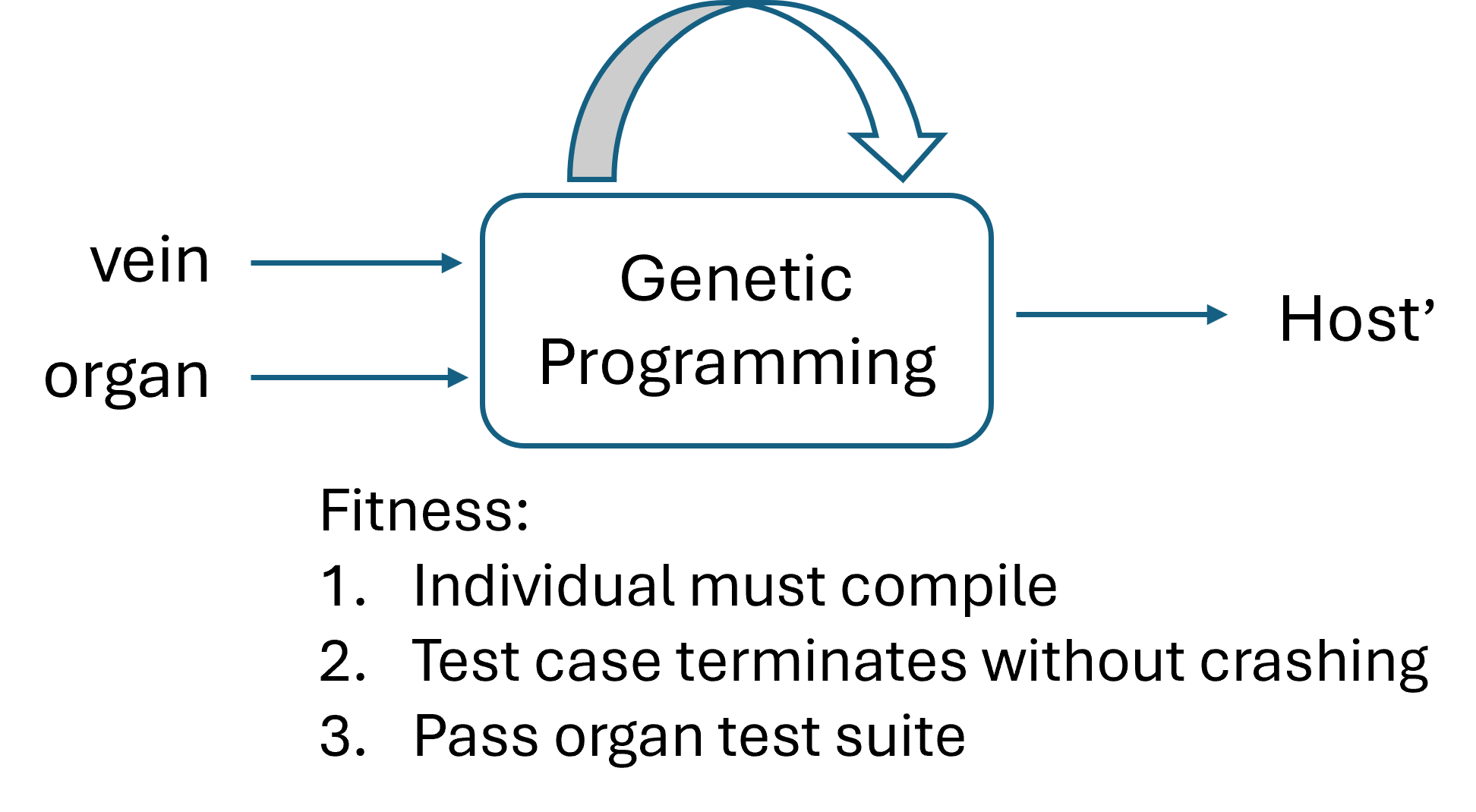}
    \caption{The organ adaptation stage of \(\mu Scalpel\) \cite{muScalpel}. The approach utilises Genetic Programming as a core transplantation algorithm. It relies on the three equally weighted fitness components.}
    \label{fig:muScalpel_organ_adaptation}
\end{figure}

Other than using GP to grow an organ from scratch or software specialisation, \citet{muScalpel, muScalpel_kate_ide} employed GP to adapt an existing organ, as implemented in their tool called \(\mu Scalpel\). Instead of inserting the entire over-organ (as described in Section \ref{sec:rq_organ_adaptation_static_analysis}) and modifying it to fit the new context, \(\mu Scalpel\) assumes that only a small number of donor statements are necessary. Hence, it incrementally adds or removes a statement for each individual in the population. 
During GP evolution, the organ's variables are matched to the host's environment.
 \(\mu Scalpel\) evaluates each individual based on three equally weighted criteria: successful compilation, termination of test execution without crashing, and passing the developer-provided organ's test suite (see Figure \ref{fig:muScalpel_organ_adaptation}).

In their subsequent work, \citet{prodScalpel} introduced \textit{ProdScalpel}, which is designed to support Software Product Line Engineering (SPLE) and is an extension of \(\mu Scalpel\)~\cite{muScalpel}. 
The approach is designed to support both domain engineering and application engineering. 
In this context, domain engineering refers to the process of extracting reusable components (over-organs) that can be adapted for specific applications, while application engineering involves the adaptation and integration of these components into a specific software product (host).

\ts{ProdScalpel}~\cite{prodScalpel} extends \(\mu Scalpel\) in four ways.
First, it supports transplanting multi-file organ and multiple organs into a single host.
Second, it addresses the problem of code redundancy that arises when transplanting multiple organs by incorporating a clone detector to avoid implanting code clones, namely NiCad~\cite{detection_and_analysis_of_near_miss_software_clones}.
Third, it manages the C preprocessor by preprocessing the organ code to remove preprocessor directives, such as IFDEFs, and its content to produce a directive-free C code.
Fourth, it solves the mutually recursive functions problem by detecting recursion using a stack mechanism.
Upon encountering a recursive function, it writes the function to a file and redirects the call to that file.

Unlike \(\mu Scalpel\)~\cite{muScalpel} that inserts an organ at a specific location and tries to adapt it, \citet{code_carbon_copy} treats an organ as a stand-alone file, where the host issues a function call from the insertion location to exercise the organ's functionality.
Hence, their organ adaptation is simply the process of variable mapping and data structure transformation between the caller and callee sites.
This is achieved by running the instrumented versions of the donor and host on the same input to capture the symbolic expressions.
This adaptation, therefore, follows a hybrid approach, where dynamic analysis is used to infer variable mappings via runtime symbolic expressions, and static analysis is subsequently applied to rewrite parameters, transform data structures, and remove irrelevant code.

\begin{figure}
    \centering
    \includegraphics[width=0.6\linewidth]{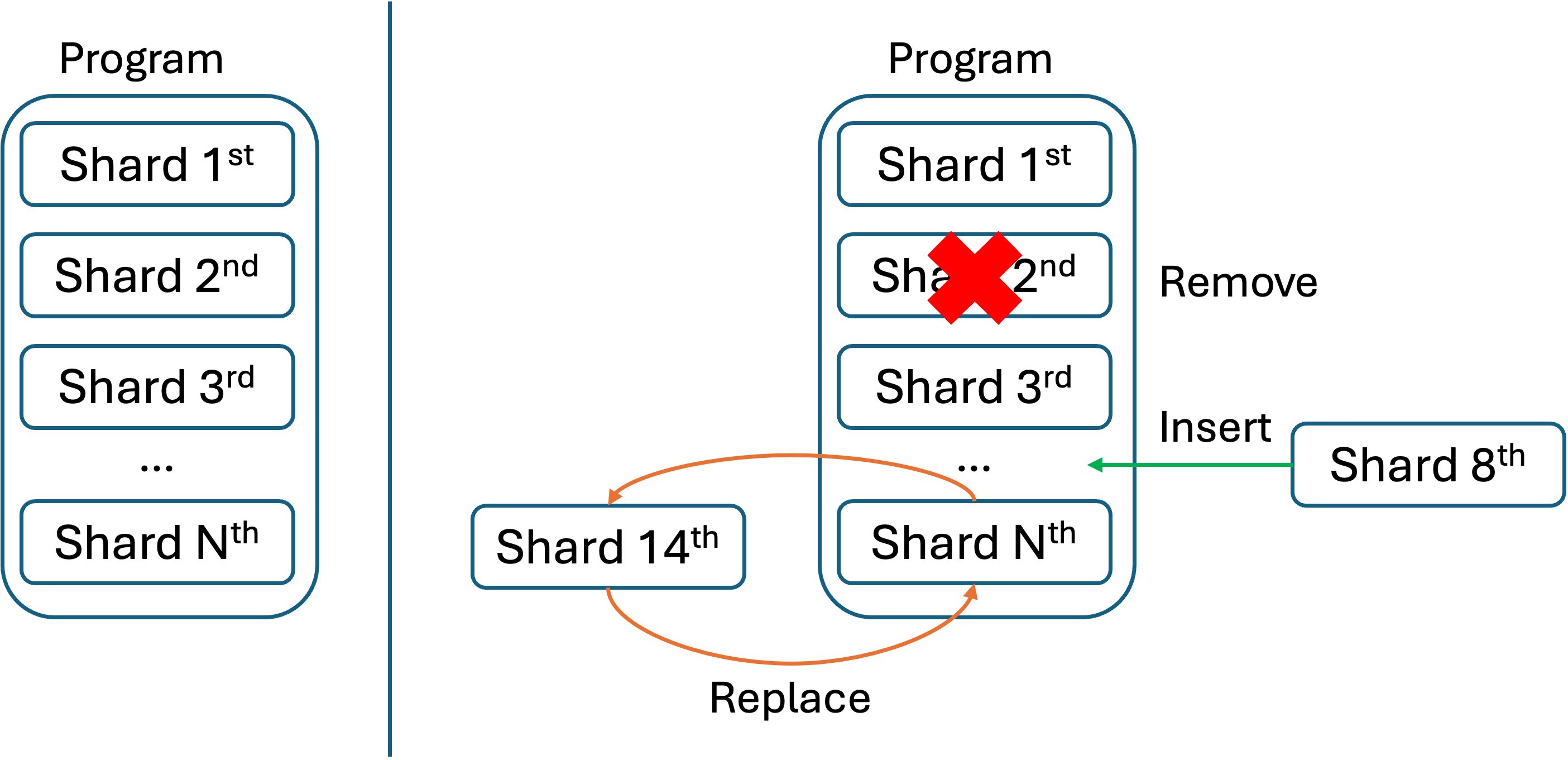}
    \caption{Program fracture and Recombination~\cite{program_fracture_and_recombination_for_efficient_automatic_code_reuse, horizontal_code_transfer_via_program_fracture_and_recombination} works by fracturing the programs into shards and classifying these shards. Then, the approach can automatically mix and match shards to obtain new software by inserting new functionality, replacing functionality with an optimised version, etc.}
    \Description{}
    \label{fig:program_fracture_organ_adaptation}
\end{figure}

Rather than adapting an organ to fit the host, \citet{program_fracture_and_recombination_for_efficient_automatic_code_reuse, horizontal_code_transfer_via_program_fracture_and_recombination} treat organ adaptation as a mix-and-match process. Given a software shard (see Section~\ref{sec:rq3_code_search}), the approach characterises its behaviour by recording input-output pairs during execution, then matches it with suitable locations in the host. Depending on the match, the approach may insert, replace, remove, or rejuvenate code in the host (see Figure \ref{fig:program_fracture_organ_adaptation}).

Apart from adapting software features, several approaches focus on migrating test cases. For example, \citet{test_transplantation_through_dynamic_test_slicing} adapts an organ simply by mocking the class and objects that were involved during the execution of the donor test but are absent in the host project.
\jp{In particular, once a donor test to be moved is identified, it is sliced, so that the organ contains a sequence of method calls on the class to which the test is to be transplanted to, using mocking for needed calls to methods absent in the host. 
This sliced test is run in the donor, to make sure it's still executable. 
Only then it is moved to the host class. 
}

Similarly, several approaches focus on migrating test cases between donor and host mobile applications. These techniques attempt to map GUI events from the donor application to corresponding events in the host application by identifying those with the highest similarity scores. 
Similarity is typically assessed by comparing widget metadata—where a widget refers to a mobile GUI element such as a button, text input field, or checkbox—based on attributes like type, text content, and screen position.

One of the earliest approaches in this domain is GUITestMigrator~\cite{test_migration_for_efficient_large-scale_assessment_of_mobile_app_coding_assignments}, which uses the Wu and Palmer (wup) semantic similarity measure~\cite{verb_semantics_and_lexical_selection} to compute similarity between preprocessed textual descriptions of GUI elements. However, GUITestMigrator often struggles to find appropriate mappings, especially when the donor and host applications differ significantly. This challenge is amplified when the donor application contains a large number of GUI events. Moreover, GUITestMigrator does not support the transplantation of test oracles.

To address some of these limitations, AppTestMigrator\cite{test_migration_between_mobile_apps_with_similar_functionality, automated_test_migration_for_mobile_apps} improves the event mapping process by using a statically generated Window Transition Graph (WTG)\cite{static_window_transition_graphs_for_android} of the host application. In this graph, nodes represent windows and edges represent transitions between them. When a direct match for a donor GUI event cannot be found in the current application state, the WTG enables forward exploration to identify future states where the target element may appear. This allows the system to guide dynamic execution toward relevant parts of the app while pruning the search space by avoiding unproductive paths. Furthermore, AppTestMigrator introduces an application-specific ontology that leverages Word2Vec~\cite{efficient_estimation_of_word_representations_in_vector_space} embeddings for more fine-grained semantic similarity, an improvement over earlier methods that relied on general-purpose lexical databases like WordNet~\cite{wordnet_a_lexical_database_for_english}.

At around the same time, CraftDroid~\cite{test_transfer_across_mobile_apps_through_semantic_mapping} was introduced with a similar objective. Like AppTestMigrator~\cite{test_migration_between_mobile_apps_with_similar_functionality}, CraftDroid constructs a UI Transition Graph (UITG) of the host application, where nodes correspond to screens and edges to GUI events. A key innovation of CraftDroid is its ability to iteratively refine the UITG at runtime by incorporating execution feedback—removing infeasible paths and integrating dynamically loaded elements. This dynamic refinement enables CraftDroid to better adapt to changes in the host application throughout the test migration process. In contrast, AppTestMigrator acknowledges that its static WTG may contain infeasible paths but does not support their dynamic correction during migration~\cite{test_migration_between_mobile_apps_with_similar_functionality}.

Despite their contributions, these three approaches — GUITestMigrator \cite{test_migration_for_efficient_large-scale_assessment_of_mobile_app_coding_assignments}, AppTestMigrator~\cite{test_migration_between_mobile_apps_with_similar_functionality,automated_test_migration_for_mobile_apps}, and CraftDroid~\cite{test_transfer_across_mobile_apps_through_semantic_mapping} — assume a one-to-one correspondence between GUI events in donor and host applications. TestMig\cite{testmig_migrating_gui_test_cases_from_ios_to_android}, however, addresses a more complex scenario: migrating test cases across platforms, specifically from iOS to Android, where direct mappings are often unavailable. To handle such many-to-many correspondences, TestMig employs a transduction technique\cite{finite-state_transducers_in_language_and_speech_processing}, governed by a predefined transduction length parameter \textit{w} and guided by a static event flow graph. Additionally, it introduces the concept of an \textit{empty UI control} to handle cases where no equivalent event exists on the target platform, thereby enabling flexible and robust n-to-m mappings across different platforms.

\ts{Similarly, TransDroid~\cite{gui_test_transfer_from_web_to_android} targets the migration of test cases from web application to Android. Rather than identifying equivalent GUI events, it applies predefined \textit{Action Transformation Rules} to reuse, merge, convert, or remove web actions, enabling platform-specific interactions to be adapted to the mobile environment.}

\ts{More recently, LLMigrate~\cite{automated_test_transfer_across_android_apps_using_large_language_models} departs from explicit event matching by using multimodal LLMs to reason about the functional intent of the donor test and dynamically select the most appropriate interaction in the host application. By combining UI layout hierarchies with screenshots, LLMigrate obtains a semantic understanding of the user interface, enabling it to migrate tests between applications within the same domain despite differences in their workflows. Furthermore, it can generate valid input values for mandatory fields in the target application that have no corresponding inputs in the source test.}

\begin{mdframed}
\textbf{
Answer to RQ3: \percentageOfApproachesUsingStaticAnalysisForOrganAdaptation{} of software transplantation approaches use static analysis for organ adaptation, while \percentageOfApproachesUsingDynamicAnalysisForOrganAdaptation{} rely on dynamic analysis. In addition, \percentageOfApproachesUsingStaticAndDynamicAnalysisForOrganAdaptation{} of approaches employ both static and dynamic analysis for organ adaptation. Among the approaches that use dynamic analysis, \percentageOfApproachesUsingGUIEventMappingForOrganAdaptation{} of approaches perform organ adaptation via GUI event mapping, while \percentageOfApproachesUsingGPForOrganAdaptation{} use Genetic Programming for tasks such as migrating features, developing new functionality from scratch, and specialising software.
}
\end{mdframed}

\subsection{RQ4: Transplantation Validation} \label{sec:rq5}
Once an organ has been transplanted into a host, software engineers should be able to verify that the host has obtained and can exercise the new functionality and that no regression fault has occurred.

\begin{figure}
    \centering
    \includegraphics[width=0.6\linewidth]{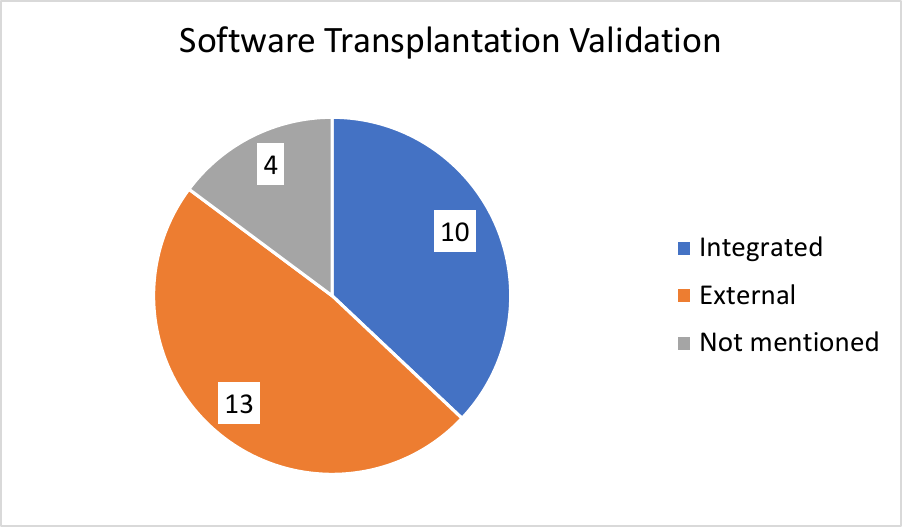}
    \caption{In total, \totalNumberOfDistinctTransplantationApproaches{} distinct software transplantation approaches were identified. Among them, \percentageOfApproachesUsingInternalValidationStep{} incorporate an integrated validation step as part of the transplantation process. In contrast, \percentageOfApproachesUsingExternalValidationStep{} rely exclusively on external validation. Finally, \percentageOfApproachesNoMentioningOfValidationStep{} of the approaches do not explicitly describe any validation mechanism.}
    \label{fig:transplantation_validation_technique}
    \Description{}
\end{figure}

Here, we present the associated verification techniques by grouping them based on the types of verification performed. We categorise the verification techniques into two main categories: (1) integrated validation steps within the transplantation approach, and (2) external validation methods that are not part of the transplantation approach. Figure~\ref{fig:transplantation_validation_technique} illustrates the distribution of how existing software transplantation approaches validate the transplantation process.

\subsubsection{Integrated Automated Validation} \label{sec:rq5:integrated_validation}
We found \numberOfApproachesUsingInternalValidationStep{} transplantation approaches that have integrated validation steps. 
We first present those that check new functionality can be correctly executed within the host, as well as check for any regressions, followed by those that only validate the transplanted functionality.

\paragraph{Validation Steps for New Functionality and Regression}

\(\mu Scalpel\)~\cite{muScalpel} relies on three validation steps.
First, it uses the existing host's regression test suite.
Second, given that an existing regression test suite might not have high coverage, authors manually augmented the test suite to increase the coverage.
Third, they manually transform the donor's acceptance test into the host context, which is used to confirm that the host has received the new functionality.
In  their subsequent work, ProdScalpel~\cite{prodScalpel} also uses the same three validation steps as \(\mu Scalpel\)~\cite{muScalpel}. 

\citet{muScalpel} performed experiments with five donors and three hosts, where \(\mu Scalpel\) automatically transplanted four out of five features, with all three validation steps passing. 
Moreover, authors conducted a case study for automatically transplanting H.264 video encoding functionality from the x264 system to the VLC media player within 26 hours, compared to an estimated 20 days of effort from human developers.
Later, this technique successfully transplanted the call graph and indentation feature from GNU \textsc{cflow} and GNU \textsc{indent} into the Kate IDE~\cite{muScalpel_kate_ide}.
The same three validation steps were used. 

For ProdScalpel~\cite{prodScalpel}, the authors conducted a user study with 20 university students, in which participants were tasked with migrating a feature into the product base. 
Authors followed the validation strategy of \citet{muScalpel}.
The study found that ProdScalpel could transplant a feature on average 4.8 times faster than the time taken by participants who successfully completed the task within the time limit. 
Furthermore, the authors carried out a case study to demonstrate ProdScalpel's capability of multi-feature transplantation, successfully transplanting features from donors both within the same domain and from different domains. 
For example, transplanting CFLOW's output and kilo's enableRawMode feature into the VI text editor as a product base.

In the context of software specialisation, \citet{justyna_specialise_minisat} validate the correctness of evolved programs using held-out test suites to ensure that the evolved version was still producing the required output.
Here the new functionality served the purpose of increasing performance, hence the test suite was not constructed to ensure new functionality was trasnplanted, but rather to ensure the overall programme produced the correct answer. 
Using this validation framework, the authors successfully specialised the MiniSAT solver for a Combinatorial Interaction Testing (CIT) problem class, achieving a 13\% reduction in runtime.

In a follow-up paper,~\citet{justyna_specialise_downstream} employed the same validation methodology and
extended their approach by specialising MiniSAT to three distinct problem domains: Combinatorial Interaction Testing, Automated Termination Analysis, and Ensemble Computation. 
They achieved runtime improvements ranging from 4\% to 36\%. 
To validate the evolved solvers, the authors used the EDACC verifier tool from the SAT 2013 competition \footnote{http://www.satcompetition.org/2013/downloads.shtml} to check the solutions generated by the final evolved solver on the held-out test suites.

For Grow and Graft~\cite{babel_pidgin}, the validations were carried out during both grow and graft stage. In the grow stage, the GP relies on the set of the fitness function, which guides each individual to the correct solution. Although the approach does not have a dedicated step to check against regression fault, its graft stage does indirectly help prevent regression fault.
This is because while inserting the organ into the host, the approach also checks for any errors, such as compilation and crash. 
During grafting, the tool automatically attempts each candidate insertion point and validates it by recompiling the host application and executing it. Any graft that causes compilation errors, runtime crashes, or incorrect functionality is rejected, ensuring that only safe insertions are retained.
 From the experiments, the approach succeeded in growing the language translation feature and grafting both the countdown and language translation features into the Pidgin chat application.

\paragraph{Only Mentioned Validation Steps For New Functionality}
For Grow and Serve~\cite{grow_django_citation}, the approach validate the new functionality in the same  in the same manner as Grow and Graft~\cite{babel_pidgin} during the grow stage. Yet, it does not have a dedicated step to check against regression fault, because the organ is not inserted into the host but rather serves as an external module. 
The authors do not mention any validation step to ensure that this host-side code does not introduce regression faults.
From the experiments, the approach successfully grew a web API service for retrieving the number of citations from Google Scholar.

For approaches that use a code search engine, users are typically asked to provide a test case to validate the transplanted organ. This is the case for tools such as Code Conjurer~\cite{code_conjurer}, Splicer~\cite{program_splicing}, and Hunter~\cite{hunter_fullPaper}. In contrast, NLP2TestableCode~\cite{NLP2TestableCode} automatically generates a JUnit test suite by analysing the returned code snippet and inferring the potential input and output types. Among these tools, Splicer offers additional flexibility: it not only accepts user-provided test cases but also allows users to supply finite automata that define valid sequences of API calls.

\citet{code_conjurer} evaluated Code Conjurer on its ability to retrieve and adapt reusable Java components from the Merobase repository. 
For each test case, a solution was considered successfully found if Code Conjurer could identify a candidate component that compiled, satisfied the interface requirements, and passed the user-provided JUnit test case. 
Using seven JUnit test cases, the tool initially identified only four solutions, reflecting the challenge of anticipating the search for complex interfaces. 
By applying the Automated Adaptation Engine (discussed in Section \ref{sec:rq3_code_search}), which uses heuristic adaptation and method-mapping to reconcile interface differences, Code Conjurer was able to identify valid solutions for all seven test cases, meaning that each candidate component compiled and passed its corresponding test.

\citet{NLP2TestableCode} evaluated NLP2TestableCode on 47 sample tasks selected from a total of 101 tasks in the original NLP2Code publication~\cite{NLP2Code}. 
They found that using lemmatisation and omitting stop words (the, an, is, etc.) during keyword processing yielded the highest number of retrieved code snippets, totalling 6,954. 
Of these, 327 code snippets were compilable without modification. 
After applying the transformation steps from NLP2TestableCode (e.g., relocating import statements to the top of the file, removing duplicate main functions, and performing line deletions in a bottom-up fashion), the total number of compilable snippets increased to 2,037. 

\citet{program_splicing} evaluated Splicer on 15 benchmark problems from diverse domains that represent common programming tasks. 
In most cases, Splicer could complete the draft program within two minutes and did so with no more than five test cases. 
They also found that the time taken by the synthesis algorithm was positively correlated with the number of holes and variables in the program. 
\ts{Next,} the authors explored the impact of using type and role matching heuristics when searching for code snippets in a search engine. 
Their findings indicated that these heuristics help reduce the search space, thereby shortening the total time Splicer needs to complete a task.
\ts{Furthermore, the authors compare Splicer against \(\mu Scalpel\)~\cite{muScalpel} on a subset of the benchmarks and found that \(\mu Scalpel\) failed to complete most of the tasks within five minutes.}

In addition,~\citet{program_splicing} conducted a user study involving 12 graduate students and six professional programmers, where each participant completed four tasks: two using Splicer and two without. The study found that Splicer was relatively easy to use, with participants needing, on average, only one invocation to successfully use the codelet returned by Splicer on the first try. Furthermore, for most programming tasks, participants completed tasks more quickly when using Splicer than when working without it. 

\citet{hunter_fullPaper} evaluated Hunter against \(S^6\)~\cite{semantics_based_code_search}, a code reuse tool with a web interface, using three different benchmarks: those from S6~\cite{semantics_based_code_search}, challenge problems from the MUSE project \footnote{http://www.darpa.mil/program/mining-and-understanding-software-enclaves}, and linked list and binary tree problems from Leetcode \footnote{https://leetcode.com}.
Out of 40 benchmarks, Hunter successfully solved all of them with an average time of 14 seconds, while S6 was able to solve only 37.5\% of the benchmarks.

Similar to Splicer~\cite{program_splicing}, Hunter also conducted a user study with three undergraduate students, 16 graduate students, and two professional programmers. 
Each participant was tasked with implementing three different programming tasks. 
Participants complete each task twice—once using Hunter to automatically synthesise the required code, and once without it. 
The results showed that using Hunter reduced the average completion time to 130 seconds with a 100\% task completion rate, while the non-Hunter method took 948 seconds on average with only 85\% completion. 
Further analysis through manual inspection and test suite verification revealed that code synthesised by Hunter was less error-prone compared to manually written code.

Nonetheless, \citet{program_fracture_and_recombination_for_efficient_automatic_code_reuse,horizontal_code_transfer_via_program_fracture_and_recombination} do not rely on test cases to validate the transplantation. 
Instead, they base validation on the matching of input-output behaviour. 
During this stage, the host's shards are executed and only those donor shards whose behaviour is semantically compatible with the host are selected. 
Given that the shard is semantically compatible, this confirms that the host will get the new functionality.

\citet{program_fracture_and_recombination_for_efficient_automatic_code_reuse} conducted two case studies.
First, they replace the triple nested loop of the matrix multiply program with BLAS library's matrix multiply.
They found that as the matrix increases in size, the faster the replaced shard runs.
In the second case study, they replaced a program with a simple dense linear solver based on Gaussian elimination without pivoting with a shard from the LAPACK linear solver library.
They found that, for larger matrix sizes, the shard replaced version runs over two times faster than the original version and exhibits superior numerical properties

Similarly,~\citet{horizontal_code_transfer_via_program_fracture_and_recombination} conducted a case study where the approach automatically replaced the sort shard from the object compatible program with the quick sort shard from the salary program.
The approach succeeded in replacing these shards.

\subsubsection{External Validation Methods}
In this section, we discuss the approaches that do not have dedicated validation steps that are part of the approach. These approaches rather rely on external methods. In total, there are \numberOfApproachesUsingExternalValidationStep{} approaches.

\citet{syntax_preserving_program_slicing} developed an automated evaluation algorithm to validate the correctness of transplantation. This is presented as Algorithm 7 in their paper~\cite{syntax_preserving_program_slicing}. The algorithm begins by executing test cases on both the donor and host versions: the donor is expected to pass the test case, while the host is expected to fail. After transplantation, the test cases are rerun on the transplanted host version, which should now pass, indicating that the new functionality has been successfully integrated. In addition to this automated check, the algorithm also produces a Git diff, allowing the author to manually inspect the changes introduced by the transplantation.

For their experimental evaluation, the authors applied their approach to 24 transplantation scenarios across different versions (git commits) of four open-source projects: scrcpy, SilverS, grep, and php. The approach successfully identified and transplanted functionality in all 24 cases.

\citet{jigsaw_fullPaper} validated the correctness of Jigsaw's organ adaptation using two complementary measures: correspondence and integration. The correspondence metric quantifies how accurately the tool identifies the correct mapping between nodes in the ASTs of the copied and pasted methods. The integration metric evaluates whether the lines of code from the copied method are inserted into the correct locations in the host. Any line that is missing or misplaced is counted as incorrect.

Using these validation criteria, Jigsaw~\cite{jigsaw_fullPaper} was evaluated in two experiments and successfully reused cloned method pairs in 12 out of 16 test cases. In a user study, participants reported that Jigsaw allowed them to focus more on higher-level tasks, such as justifying and confirming the rationale for reusing a particular method, rather than dealing with low-level integration details.

For machine learning approaches, AdaptivePaste~\cite{adaptivePaste} does not have a dedicated validation method integrated within the tool.
In the experiments, the authors define two evaluation metrics: `exact match accuracy' and `ValidPaste accuracy'.
Exact match accuracy is a ratio of the number of code snippets de-anonymised correctly divided by the total, while ValidPaste accuracy is a ratio of predicted code snippets with all bound variables de-anonymised correctly to the total.

From the experiments, \citet{adaptivePaste} evaluated two variants of AdaptivePaste (uni-decoder and parallel-decoder model) against three baseline models, namely Masked Language Modeling~\cite{roberta_a_robustly_optimized}, DOBF model~\cite{dobf_a_deobfuscation_pre_training}, and CodeT5 model~\cite{codeT5_identifier_aware}.
Among the two variants of AdaptivePaste, the uni-decoder model performed the best in terms of the accuracy and semantic validity of the adapted code.
Furthermore,~\citet{adaptivePaste} conducted a user study with 20 developers with a real-world application.
The participants were divided into two groups, where each group worked on different tasks.
The first group manually predicted anonymised identifier names, given the surrounding code context.
The second group used the AdaptivePaste tool to adapt the code.
The results of the study showed that AdaptivePaste was able to generate accurate adaptations, whereas manual adaptation proved challenging, especially for developers who are unfamiliar with the code's context or lack of knowledge about the specific API.
However, AdaptivePaste can seamlessly handle these adaptations, even in such cases.

Another machine learning approach, \citet{smartPaste} evaluated the model's correctness by whether the predicted variables matched the original variable names before they were replaced with placeholders. The projects were split into training, validation, and test sets, with the test data further divided into seen and unseen projects, allowing the authors to assess both performance and generalisation. Model performance was measured using multiple quantitative metrics. For instance, accuracy is the percentage of correct predictions, mean reciprocal rank (i.e., MRR) is the mean reciprocal rank of each prediction, and type match is the percentage of suggestions of the correct type.

The experimental results show that the proposed models (e.g., AVGG, GRUG, GRUD, and HD) significantly outperform the baseline LOC model, demonstrating that they learn from usage context rather than relying solely on type information. The models were able to distinguish between variables of different types, and removing type information reduced performance by approximately 8\%. In terms of generalisation, models still managed to achieve a good performance on unseen projects, despite a slightly lower performance compared to the seen projects.

Lastly, \citet{towards_automatic_client-side_feature_reuse, DBLP:phd/se/Maras14} validated their client-side web UI feature reuse approach using functional testing and quantitative case-study metrics. To verify the correctness of the host application after feature transplantation, the authors employed Selenium-based tests to check that the presentational and behavioural characteristics of the reused features were preserved. The evaluation was conducted using Firecrow, a Firefox plugin implementing the proposed reuse algorithms, and several metrics were collected for each case study, including the size of the source and target applications, the number of automated modifications required to resolve HTML, CSS, and JavaScript conflicts, the execution time of the reuse process, and whether the reuse attempt was ultimately successful.

Using this validation setup, \citet{towards_automatic_client-side_feature_reuse} conducted four case studies involving feature reuse across web applications developed with different JavaScript libraries. Two applications were built using jQuery and two using MooTools, enabling both cross-library reuse (jQuery to MooTools and vice versa) and reuse between applications using the same library. The approach successfully reused features in three out of the four case studies, effectively resolving conflicts related to HTML attributes, CSS overrides, and library-specific method usage. The single failure case occurred due to complex prototype extensions in MooTools that the tool could detect but not automatically repair.

\paragraph{Manual Validation}
In addition, \numberOfApproachesManualExternalValidationStep{} studies manually validate the approach. From the pragmatic reuse domain, \citet{procrustes} evaluated the Procrustes tool in two stages.
First, the authors evaluated the approaches by manually and using the Procrustes tool to perform the reuse task.
They measure the required effort by counting the number of compilation errors and the minimum number of edits required to successfully complete the reuse task.
Second, the authors conducted a user study where they used two test suites (e.g., Metrics Eclipse plug-in and Azureus client application) to validate the reuse task.
In the user study, the authors used various metrics such as the number of compilation errors, time spent on completing the task, etc. 
More importantly, they used the test suites to validate the success of the task.

Using this validation framework, Procrustes~\cite{procrustes} was evaluated on two non-trivial reuse tasks: the line of code calculator from the Metrics Eclipse plug-in \footnote{http://metrics.sf.net/} and the network throughput visualisation view from the Azureus client application \footnote{http://azureus.sf.net/}. For the Metrics task, the tool reduced compilation errors by 95.2\% and required 96.7\% fewer edits. In the Azureus task, compilation errors decreased by 65.6\% and edits by 84\%. In a controlled study with eight developers, all tasks performed with Procrustes were successful, compared to only 50\% success when performed manually. Time savings were substantial: the Metrics task averaged 19 minutes with Procrustes versus 47 minutes manually, and the Azureus task averaged 10 minutes versus 40 minutes.

Gilligan~\cite{gilligan} is an Eclipse plug-in that provides a Validation and Error view to assist users during code reuse. The Validation view displays a list of components that still need to be triaged with a reuse decision (e.g., accept or reject). The Error view displays errors every time the user enacts the pragmatic reuse plan. However, the responsibility for determining whether the reuse is correct and complete remains with the user, who must repeated inspect the provide GUI views and modify the reuse accordingly.

\citet{gilligan} conducted an evaluation of Gilligan in two stages with 16 participants.
In the first group, participants used Gilligan in stage one and a standard IDE tool in stage two, while the second group used a standard IDE tool in stage one and Gilligan in stage two.
During the first stage, participants were tasked with reusing two features: the QIF parser functionality from jGnash \footnote{http://jgnash.sf.net/} and the related artists list feature from aTune \footnote{http://atunes.sf.net/}.
The second phase focused on measuring the time it took the software engineers to realise the reuse task.
The results of the experiment showed that Gilligan enabled developers to complete reuse tasks 38\% faster.

For the symbolic expression approach,~\citet{code_carbon_copy} experimented on CCC with three donors and five hosts.
From the experiments, CCC successfully transplanted seven out of eight transplantation scenarios, whereas the last scenario failed due to the fact that the approach did not support that data translation between the organ and the host.
The authors manually validate the result by considering the transfer to be successful if and only if the their manual analysis of the transfer shows that it can correctly produce output for all input files.
Yet, they also use other metrics, such as the wall clock time spent by the approach, the total transferred lines of code, etc.

In terms of GUI test reuse, \citet{test_transfer_across_mobile_apps_through_semantic_mapping} evaluated CRAFTDROID with 25 Andriod applications, which come from five different categories (e.g., browser, to-do list, shopping, mail client, and tip calculator).
For each category, the authors identified two main functionalities that will be transplanted between applications within the same category.
CRAFTDROID will transfer the test case to four other applications within the same category.
After each transfer, the authors manually examined the transferred test.
From the experiment, CRAFTDROID, on average, successfully transfer 74.5\% of the attempts.
There is a negative correlation between the test length and the transfer success rate.
In addition, they found that, on average, test transfer takes less than 1.5 hours.

\citet{test_migration_for_efficient_large-scale_assessment_of_mobile_app_coding_assignments} evaluated GUITestMigrator using three sets of Android mobile applications developed by students in an online course. 
These applications included a bowling alley management system, a grocery list app, and a coffee cart reward and payment system.
The authors first randomly selected and collected test cases written in the Espresso testing framework from six applications. 
These test cases were selected from the applications that the authors can successfully run and execute.
Next, they used GUITestMigrator to migrate each of these collected test cases to the corresponding host applications, which were randomly selected from the same set of applications.
During the experiment, the authors manually inspected the GUI events migrated from the donor applications to the host applications.

From the experiments, \citet{test_migration_for_efficient_large-scale_assessment_of_mobile_app_coding_assignments} found that GUITestMigrator accurately migrated 88\% of the GUI events.
Regarding the length of the GUI event sequence, the authors discovered that GUITestMigrator successfully migrated the longest possible sequence for 60\% of the host applications.
Through manual inspection, the authors identified and classified the root causes of unsuccessful test case migrations into six categories: (1) the use of different GUI events to perform a task, (2) getting lost during dynamic exploration, (3) missing or unmatched GUI events, (4) the use of widgets with unusual labels, (5) the use of predefined lists, (6) application-specific issues that could not be classified.

In their subsequent work, \citet{test_migration_between_mobile_apps_with_similar_functionality} evaluated AppTestMigrator using 16 Android applications from four randomly selected categories: shopping list, note taking, expense tracking, and weather apps.
Since AppTestMigrator was built upon GUITestMigrator, the authors also compared the performance of these two approaches.
During the experiment, the authors manually examined the migrated test cases.
With AppTestMigrator, they checked both the test events and the oracles.
Conversely, they only check for migrated test events from GUITestMigrator.

From the experiments, \citet{test_migration_between_mobile_apps_with_similar_functionality} found that, on average, AppTestMigrator completely migrated 48\% and partially migrated 34\% of test cases. 
In terms of test oracle migration, AppTestMigrator correctly matched 47\% of the assertions in the test cases, on average. 
Compared to GUITestMigrator, AppTestMigrator completely migrated 16\% more test cases but partially migrated 5\% fewer test cases.

\citet{testmig_migrating_gui_test_cases_from_ios_to_android} evaluated TestMig with five applications (e.g., 2048, SimpleNote, Wikipedia, Wire and WordPress) that have both iOS and Andriod version.
The authors manually examined each generated Android test case by comparing the behaviour against the original iOS test case.
They defined three metrics: Test Migration Rate (TMR), Partial Migration Rate (PMR), and Event Migration Rate (EMR).
From the experiments, they found that TestMig, on average, correctly converted 80.2\% of the iOS UI test cases to Android test cases.
The migrated test cases achieve a similar statement coverage compare to the original iOS test cases with 59.7\% and 60.4\% for Android and iOS, respectively.
Regarding transduction length, the study showed that increasing transduction length from one to five generally improve all three migration rates.
Yet, the rate stabilised once the transduction length reaches a value of three.
This suggests that one iOS event is typically mapped to at most three Andriod events.
Moreover, the authors classified 26 test cases that failed to fully migrated into three categories: missing feature in Android versions, an approach's inability to migrate the iOS system event to Android, and multiple similar following GUI events.

\ts{\citet{gui_test_transfer_from_web_to_android} evaluated TransDroid using 10 pairs of real-world web and Android applications (e.g., BuzzFeed, DokuWiki, Etsy, Fox News, GitLab, Groupon, Hacker News, OwnCloud, Wikipedia, and WordPress).
In their evaluation, the authors manually constructed corresponding Android tests as ground truth for assessing transfer success and manually examined the generated tests.
From the experiments, they found that 77\% of the attempted transfers were successful.
They also found that test length was not a key factor affecting the effectiveness of cross-platform test transfer, which contrasts with the earlier finding of \citet{test_transfer_across_mobile_apps_through_semantic_mapping} for intra-platform mobile test transfer.
To assess effort reduction, the authors computed the Levenshtein distance between each migrated test and its corresponding ground-truth test.
They found that the migrated tests required only 0.98 manual edits on average to match the ground-truth test.
This corresponds to an average effort reduction of 82\%, suggesting that TransDroid can substantially reduce the manual effort required to create Android GUI tests.
Finally, TransDroid required less than six minutes on average to complete a single test transfer.
}

\ts{\citet{automated_test_transfer_across_android_apps_using_large_language_models} evaluated LLMigrate using 19 Android applications from five categories (e.g., browser, to-do list, shopping, mail client, and tip calculator). The authors compared LLMigrate against three existing test-transfer approaches, namely CraftDroid~\cite{test_transfer_across_mobile_apps_through_semantic_mapping}, TREADroid~\cite{enhancing_test_reuse_with_gui_events_deduplication_and_adaptive_semantic_matching}, and TEMdroid~\cite{learning_based_widget_matching_for_migrating_gui_test_cases}. Given that there can be more than one correct transferred test, the authors manually inspected all transferred tests from all techniques. The experimental results show that LLMigrate achieved a 97.5\% successful transfer rate and reduced the manual effort required to write tests by 91\%. In addition, LLMigrate completed each transfer in 247 seconds on average.}

\ts{Furthermore, \citet{automated_test_transfer_across_android_apps_using_large_language_models} evaluated LLMigrate on additional recent real-world applications, reporting an average values of 98\% for success rate, 97\% for reduction score, and 238 seconds for transfer time. Finally, to assess generalisability beyond data likely seen by the underlying LLMs, they evaluated LLMigrate with 15 Android applications from five new app categories (e.g., EMI Calculator, AI Chatbots, Movies, Note, and Messenger), achieving a 93\% successful transfer rate and an 88\% reduction score.}

\subsubsection{No Mentioning of Validation Steps}
Lastly, \numberOfApproachesNoMentioningOfValidationStep{} approaches did not provide any details about the validation steps.
First, \citet{a_study_on_code_transplantation_technique} experimented with two calculator programs from GitHub, which they called Calculator1 and Calculator2. The Calculator1 program has a feature and a beautiful GUI. Conversely, the Calculator2 program does not possess any calculating ability and has a poor UI. In the experiment, the authors transplant the calculating and beautiful GUI feature from Calculator1 into Calculator2. Unfortunately, they did not provide any details about the experimental results.

Second, \citet{a_study_on_tree-based_code_extraction} conducted a case study on two open-source notepad programs from GitHub. Notepad1 program is a complete software equipped with the necessary features, while Notepad2 program is missing two features: an open file feature and can only save files to the current directory. So, their case study focuses on transplanting these missing features from Nodepad1 into Nodepad2. Similarly, they did not provide any details about the experiment.

Third, in the case of \citet{reusing_web_application_user-interface_controls}, the process used to validate the transplantations was unclear. While the authors stated that they evaluated their approach by extracting UI controls from 35 web pages, they did not explain how they assessed the success of these extractions. Furthermore, they provided no discussion regarding the reuse of the extracted components in the host web pages. Although we found a YouTube video\footnote{\url{https://www.youtube.com/watch?v=-7Ead_Zbqq4}} demonstrating Firecrow, its connection to the publication was not explicitly established.

Fourth, \citet{test_transplantation_through_dynamic_test_slicing} presented an approach for transplanting a test case via dynamic slicing as a proof-of-concept.
They have not conducted experiments to validate their approach. 

\begin{mdframed}
\textbf{
Answer to RQ4: \numberOfApproachesUsingInternalValidationStep{} approaches (\percentageOfApproachesUsingInternalValidationStep{}) have a validation step integrated within the approach, while \numberOfApproachesUsingExternalValidationStep{} approaches (\percentageOfApproachesUsingExternalValidationStep{}) rely on external validation steps.
Among those relying on external validation, \numberOfApproachesManualExternalValidationStep{} approaches (\percentageOfApproachesManualExternalValidationStep{}) require humans to manually validate the results.
The other \numberOfApproachesNoMentioningOfValidationStep{} approaches (\percentageOfApproachesNoMentioningOfValidationStep{}) do not mention how internal/external validation was undertaken.
}
\end{mdframed}

\subsection{RQ5: Automation of Software Transplantation}
This research question aims to understand to what extent has the automation been adopted or integrated as part of software transplantation. 
We look at the automation of software transplantation at three stages: organ location, organ extraction and adaptation, and validation.

\subsubsection{Organ Location}
The first step of software transplantation is the location of a suitable organ. 
From our literature review, we found that all but \numberOfApproachesNotRequireOrganLocation{} proposed approaches required the user to supply an organ as the first step. 
Although some simply ask for a search query, others require more extensive manual effort from the user.
For more in-depth detail, we refer readers to Section \ref{sec:rq_organ_location}.

Organ location can be a manual task, like in the pragmatic reuse domain~\cite{procrustes, gilligan, semi_automating_pragmatic_reuse_tasks}. 
With this approach, software engineers need to label all components with a reuse decision.
For a greater degree of automation during this stage, some approaches only ask a user to provide the organ entry points~\cite{muScalpel} or search query~\cite{code_conjurer}.

Among \totalNumberOfDistinctTransplantationApproaches{} approaches, we identified \numberOfApproachesNotRequireOrganLocation{} that do not include an organ location stage, namely Grow and Graft \cite{babel_pidgin}, Grow and Serve \cite{grow_django_citation} and software specialisation \cite{justyna_specialise_downstream, justyna_specialise_minisat}.
However, we note here that while growing an organ from scratch may seem straightforward and requires no direct human intervention in terms of location, the process of defining the organ's specification and designing the heuristic algorithm's fitness function is manual.
This stage demands significant human effort to carefully identify and define the necessary parameters for the organ's creation.

\subsubsection{Organ Extraction and Adaptation}
Once an approach reaches the organ extraction and adaptation stage, we start to see more automation. 
In the pragmatic reuse domain, once the user supplies the pragmatic reuse plan, the approach automatically locates and moves the relevant code snippets from the donor into the host.
 
On the other hand, \citet{smartPaste}'s approach automatically adapts the copied code snippet into the new location via deep neural models, once the user selects the target code snippet and pastes at the new location.

Yet, there are two approaches that still require user input during the automated process of organ adaptation to resolve conflicts, namely Jigsaw \cite{jigsaw_2pages, jigsaw_fullPaper} and the approach by \citet{reusing_web_application_user-interface_controls}.

Although~\citet{a_study_on_tree-based_code_extraction, a_study_on_code_transplantation_technique} proposed a technique that relies on program slicing and AST manipulation, they did not propose any tool to support their approaches.
Moreover, they did not provide much detail about how the approach works.
Hence, it was not clear to what extent their approach is automated.

\subsubsection{Validation}
This subsection explores the automation in the validation step of software transplantation. 
From RQ5, we observe that the validation of software transplantation approaches can be categorised into two groups: those integrated within the approaches themselves and those using external validation methods.
\ts{From the literature, \numberOfApproachesUsingInternalValidationStep{} out of the \totalNumberOfDistinctTransplantationApproaches{} proposed approaches have integrated and automated their validation steps.}

For example, \(\mu Scalpel\)~\cite{muScalpel} automatically tests each individual in the population to evaluate its fitness function.
Toward the end of the process, it also validates the best individual against an augmented version of the test suite and checks for any regression faults.

Similarly, techniques that rely on code search engines often either request the user to supply test cases as a search query or automatically generate test cases based on existing information.
For example, Code Conjurer~\cite{code_conjurer} asks the user to provide a test suite as a query. 
Once the code is retrieved from the database, the user can request the tool to test the retrieved code with the previously provided test suite.
In contrast, NLP2TestableCode~\cite{NLP2TestableCode} automatically generates a simple JUnit test suite by inferring the input and output types of the target code snippet and transforming it into a testable function.

Nonetheless, \numberOfApproachesManualExternalValidationStep{} approaches still require users to perform manual validation.
For instance, \citet{code_carbon_copy} manually verifies that the host can correctly produce output for all input files.
\citet{test_transfer_across_mobile_apps_through_semantic_mapping} manually examined the generated tests and their execution in order to calculate the success rate, precision and recall.
Similarly, \citet{testmig_migrating_gui_test_cases_from_ios_to_android} also manually inspected the generated tests and only considered a migration to be successful if and only if all iOS's GUI events were correctly mapped to Android's.

\begin{mdframed}
\textbf{Answer to RQ5: Automation in all software transplantation approaches starts once an approach receives the organ location, which could be an entry point in code, returned code snippet(s), or reuse plan.
Yet, \numberOfApproachesRequireUserInterventionDuringAdaptation{} approaches (\percentageOfApproahcesRequireUserInterventionDuringAdaptation{}) required the user to help resolve conflicts.
In terms of validation, \percentageOfApproachesUsingInternalValidationStep{} of the proposed approaches employ integrated validation steps, which are automated and executed as part of the transplantation process itself.
}
\end{mdframed}

\subsection{RQ6: Benchmarking}

Here we report on the types of software used in empirical work on software transplantation. 
We comment on whether they could be used as benchmarks for future studies. 

We note here that this section includes all publications that mention particular software, regardless of whether it was used in an experimental setting, or just used to calculate statistical data (such as determining how many components can be extracted from a specific \texttt{git} commit) or in a proof-of-concepts, without implementation.
In total, we identified \numberOfPapersNameSoftwareForTransplantationPurpose{} papers that name software for transplantation purposes. 

In Section \ref{sec:rq_benchmarking_empirical_benchmark}, we separately consider software used in empirical transplantation studies only. 
These demonstrate the feasibility of the proposed transplantation techniques in practice. 
We consider whether software used in those studies could be used as benchmarks in future work.

\subsubsection{Software Functionality of Interest}

We classify software used in transplantation literature into two categories: those offering new software features and those focusing on test transplantation.
Among the papers that provide software artefacts, \numberOfPapersDiscussSoftwareFeature{} papers (\percentageOfPapersDiscussSoftwareFeature) focus on reusing features and \numberOfPapersDiscussTestCase{} papers (\percentageOfPapersDiscussTestCase{}) focus on reusing tests \ts{\notePercentageRounding{}}.

\paragraph{Reusing Software Feature}
Among approaches that focus on software feature reuse, prior work has transplanted a diverse set of software artifacts drawn from existing systems. From a web reuse domain, \citet{reusing_web_application_user-interface_controls, towards_automatic_client-side_feature_reuse, DBLP:phd/se/Maras14} worked on transplanting user interface control from a donor website into a host website. However, these studies do not specify the exact name of the transplanted organ nor clearly state the donor and host websites involved.

Apart from web reuse, \citet{muScalpel} transplanted the H.264 video encoding functionality from x264 into VLC media player, while \citet{code_carbon_copy} transplant image rotation functionality from mtPaint graphics editor into cwebp image compressor. In the context of software product lines, \citet{prodScalpel} transplanted several features from different text editors into VI and VIM editors. 
\citet{justyna_specialise_downstream, justyna_specialise_minisat} transplanted code from different versions of MiniSAT Boolean satisfiability solver into the original. 

In contrast, \citet{syntax_preserving_program_slicing} investigate functionality transplantation between different versions of the same software system, including scrcpy, which allows mirroring of Android devices, as well as unix search utility of grep, and others. 

Furthermore, certain transplantation approaches support transfer at a smaller granularity that resembles copy-and-paste reuse. 
For example, Code Conjurer~\cite{code_conjurer}, Hunter~\cite{hunter, hunter_fullPaper}, Jigsaw~\cite{jigsaw_2pages, jigsaw_fullPaper}, and NLP2TestableCode~\cite{NLP2TestableCode} are IDE plugins that retrieve reusable code snippets from program databases and adapt them to the host context. 
Although these approaches operate at a finer granularity than traditional software transplantation, the retrieved artifacts still implement specific functionality of interest (i.e., organ). 
In practice, the reused components often correspond to small but self-contained tasks, such as utility functions (e.g., matrix multiplication, thread creation, or palindrome detection), as well as snippets extracted from real-world systems such as the Eclipse Java Development Tools (JDT) and Eclipse Annotation Processing Tool (APT).

Similarly, in AdaptivePaste~\cite{adaptivePaste}, the transplanted organ is a code snippet. Snippets are extracted from a project and adapted back into the same project after anonymisation, simulating the integration of externally sourced code. These snippets are chosen based on structural similarity to code found on Stack Overflow. The study uses five large open-source Python systems, such as a RethinkDB, a database for building realtime web applications.

Lastly, some researchers demonstrate the feasibility of their approach with the artificial benchmark. \citet{horizontal_code_transfer_via_program_fracture_and_recombination} transplant a Quicksort algorithm from a salary-processing application into an object-compaction application, replacing an existing Bubble Sort implementation. \ts{\citet{babel_pidgin} grow language translation organ from scratch and later transplant it into the Pidgin chat application.}

\paragraph{Reusing Tests}
Another line of research explores the reuse and transplantation of test cases across applications or development ecosystems. 
One common scenario is the migration of GUI test cases between mobile applications with similar functionality. 
For example, \citet{test_transfer_across_mobile_apps_through_semantic_mapping} used CraftDroid to transplant a Remove Task GUI test from a To-Do List application to a Shopping List application by semantically mapping user interface elements between the two apps. 
Similarly, \citet{testmig_migrating_gui_test_cases_from_ios_to_android} used TestMig to transplant GUI test cases from iOS to Android for Wikipedia and others. 
\ts{In constrast, \citet{gui_test_transfer_from_web_to_android} used TransDroid to transplant GUI test cases from web to Android for Hacker News and others.}

Another setting where test reuse has been explored is within fork ecosystems, where independently evolving forks often contain unshared tests that target similar functionality. 
\citet{semi-automated_test-case_propagation_in_fork_ecosystem} found that 75\% of ecosystems contained non-shared test cases that targeted similar functionality in other forks, suggesting a high potential for reuse. 
\citet{to_share_or_not_to_share} found that 48\% of newly added test cases were missing in at least one fork and that 24\% of those missing tests targeted matching units under test, making them viable candidates for automated propagation.

\subsubsection{Benchmarks} \label{sec:rq_benchmarking_empirical_benchmark}

\begin{figure}
    \centering
    \includegraphics[width=0.6\linewidth]{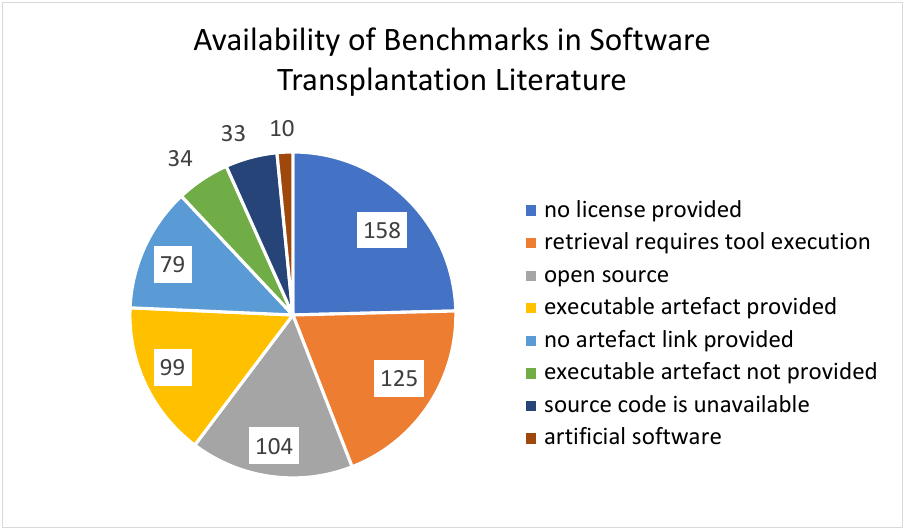}
    \caption{We found a total of \numberOfBenchmarksTotal{} benchmarks, \numberOfBenchmarksNoLicenseProvided{} of which do not have a permissive license, followed by \ts{\numberOfBenchmarksRetrievalRequiresToolExecution{} to which an artefact requires tool execution to retrieve.} Only \numberOfBenchmarksArtificialSoftware{} benchmarks are artificial, i.e., hand-crafted to illustrate a given approach.}
    \Description{}
    \label{fig:availability_of_benchmark_from_software_transplantation_literatures}
\end{figure}

Next, we looked into which software has been used for transplantation in empirical studies. 
These show the feasibility of the approach in practice. 
We classified each software artefact into \ts{eight} categories: ``no artefact link provided'', \ts{``retrieval requires tool execution'',} ``source code is unavailable'', \ts{``executable artefact not provided'',} ``no license provided'', ``open source'', \ts{``executable artefact provided'',} and ``artificial software''.

Figure \ref{fig:availability_of_benchmark_from_software_transplantation_literatures} presents the distribution of software artefacts across categories. 
We observed that \percentageOfBenchmarksOpenSource{} of all artefacts are open source.
Although we did not find any closed-source software, we came across \numberOfBenchmarksSourceCodeIsUnavailable{} unavailable source code.
For example, the provided artefact link is incorrect and redirects us to the ``404 page not found'' page.
Similarly, \numberOfBenchmarksNoLicenseProvided{} artefacts are missing a license that allows for software reuse.

Additionally, we identified \numberOfBenchmarksArtificialSoftware{}  systems in the literature, i.e., created by the authors to illustrate their ideas. Six of these serve as hosts, three are organs, and one serves as a donor.

\begin{figure}
    \centering
    \includegraphics[width=0.6\linewidth]{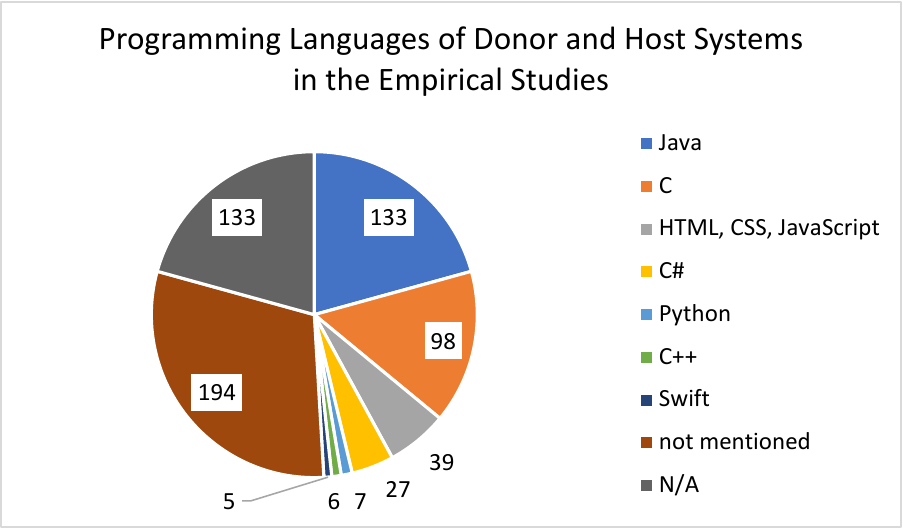}
    \caption{\ts{The distribution of programming languages for the donor and host systems that were used in the empirical studies.}}
    \label{fig:programming_language_for_donor_and_host_systems_in_empirical_studies}
    \Description{}
\end{figure}

\ts{Figure \ref{fig:programming_language_for_donor_and_host_systems_in_empirical_studies} shows the programming languages of donor and host systems used in empirical software transplantation studies.
Java (\percentageOfBenchmarkLanguageJava{}) and C (\percentageOfBenchmarkLanguageC{}) are the dominant languages.
However, for \percentageOfBenchmarkLanguageNotMentioned{} of the benchmarks, the programming language of the donor or host system is not explicitly reported.
This limits the extent to which we can characterise the implementation ecosystems used in empirical studies.
Among the explicitly reported languages, the results suggest that empirical software transplantation studies have mostly focused on a small number of ecosystems, particularly Java- and C-based systems.}

\begin{figure}
    \centering
    \includegraphics[width=0.6\linewidth]{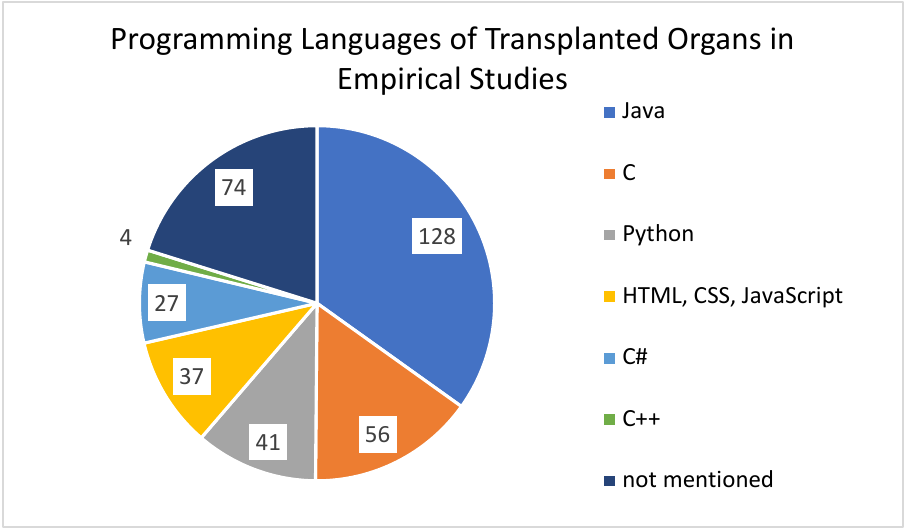}
    \caption{The distribution of programming languages for the \ts{transplanted} organs that were used in the empirical studies.}
    \label{fig:programming_language_for_transplanted_organ_in_empirical_studies}
    \Description{}
\end{figure}

\ts{We observe a broadly similar trend for the programming languages of transplanted organs (see Figure \ref{fig:programming_language_for_transplanted_organ_in_empirical_studies}).
Java is again the most frequently used language (\percentageOfOrganLanguageJava{}), followed by C (\percentageOfOrganLanguageC{}).
However, the organ-language distribution is more explicit than the donor/host distribution, with fewer cases where the programming language is not mentioned (\percentageOfOrganLanguageNotMentioned{}).
Moreover, we can see that there is little interest in programming languages, such as JavaScript and Python, despite nowadays they are widely used for modern development tasks.}

We hypothesise that the limited emphasis on Python in software transplantation research can be attributed to two main factors. First, early studies in this field primarily focused on languages such as C and Java. Consequently, many tools and techniques were designed and implemented specifically for these languages. This has encouraged subsequent researchers to continue using them, as doing so facilitates direct comparison with existing work. Such persistence reflects the concept of social inertia \cite{wikipedia_social_inertia}, where established practices persist over time despite the availability of newer alternatives. In the context of software transplantation research, this can manifest as a reluctance to explore new programming languages.

Second, Python's dynamic typing poses challenges for the transplantation process. Prior to applying transplantation algorithms, the code must be analysed to identify potential organs and their dependencies. Many existing approaches rely on static analysis for this task. However, because Python does not provide type information without code execution, static analysis becomes significantly more complex \cite{risky_dynamic_typing_related_practices_in_python_an_empirical_study}. This makes it harder to accurately detect organs and dependencies. In contrast, statically typed languages such as C and Java support more straightforward static reasoning.

\begin{figure}
    \centering
    \includegraphics[width=0.6\linewidth]{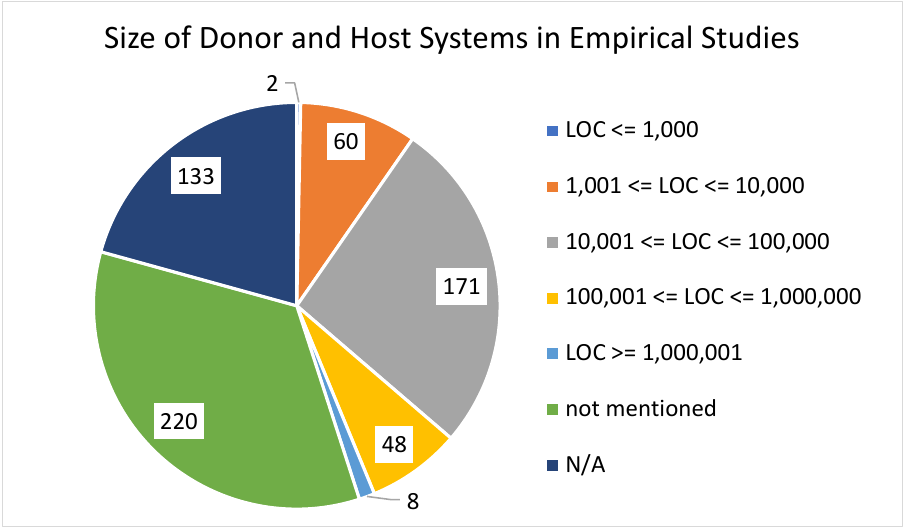}
    \caption{The distribution of benchmarks that were used in the empirical studies, in terms of lines of code (LOC). \ts{Thanatad updated this figure.}}
    \label{fig:size_of_donor_and_host_systems_in_empirical_studies}
    \Description{}
\end{figure}

In terms of software size, we classify each benchmark into five groups: from less than 1,000 to greater than 1 million LOC \ts{(see Figure \ref{fig:size_of_donor_and_host_systems_in_empirical_studies}).}
Benchmarks with less than 1,000 LOC include Mytar from \citet{muScalpel}, and Kilo from \citet{prodScalpel}.
Benchmarks from 10,000 LOC to more than 1 million LOC, including software such as Pidgin, VLC, and PhP Interpreter.

Furthermore, we observed that some benchmarks are very well-known among researchers in this domain.
We found that \numberOfBenchmarksUsedAcrossDifferentStudies{} benchmarks were used in \numberOfStudiesUsingRepeatedBenchmarks{} different studies.
\ts{Examples include} Azureus project\footnote{http://azureus.sf.net}, GNU Cflow\footnote{https://www.gnu.org/software/cflow/}, MiniSAT\footnote{http://minisat.se/MiniSat.html}, MyTar\footnote{https://github.com/spektom/mytar}, Pidgin\footnote{https://www.pidgin.im}, and VLC\footnote{http://www.videolan.org/}.
\numberOfBenchmarksUsedAsBothDonorAndHost{} benchmarks were used as both donor and host in the transplantation process.

\begin{mdframed}
\textbf{
Answer to RQ6: The most commonly used software artefacts in the literature on software transplantation are software features (\percentageOfPapersDiscussSoftwareFeature{}), followed by test cases (\percentageOfPapersDiscussTestCase{}).
\percentageOfBenchmarksOpenSource{} of benchmarks used in empirical studies are open source.
\ts{Java is the most frequently used programming language for transplanted organs (\percentageOfOrganLanguageJava{}).}
\numberOfBenchmarksUsedAcrossDifferentStudies{} benchmarks were reused across \numberOfStudiesUsingRepeatedBenchmarks{} different studies, while \numberOfBenchmarksUsedAsBothDonorAndHost{} programmes served as both donor and host in the transplantation processes.
}
\end{mdframed}

\subsection{RQ7: Industrial Adoption of Software Transplantation}
Despite an increase in publications since Harman's keynote paper on software transplantation in 2013, we found no instance of adoption of the approach in practice. 
Nonetheless, case studies with human participants (e.g., professional software engineers and university students) have been performed. 
Unfortunately, only a minority of publications include such studies. 
For instance,~\citet{jigsaw_fullPaper} evaluated Jigsaw with two software engineers, where they worked on three unique tasks, which are different from each other. 
Participants were asked to use Jigsaw to perform small-scale reuse tasks.

In the machine learning domain,~\citet{adaptivePaste} conducted a user study on AdaptivePaste with 20 software engineers, whose experience ranged from 2 to 25 years. 
One group of participants was tasked with predicting anonymised identifier names based on the surrounding code context, while a second group used AdaptivePaste to adapt the code.

In the code search domain,~\citet{program_splicing} experimented using Splicer with six software engineers and 12 students. 
Each participant completed four tasks: two without Splicer and two using Splicer. 
Similarly,~\citet{hunter, hunter_fullPaper} conducted a user study involving 19 students and two professional software engineers, where participants implemented three Java programming tasks. 
They were asked to repeat each task twice by using Hunter to automatically synthesise the code and manually implement it.

In the pragmatic reuse domain,~\citet{gilligan} ran a user study with seven professional software engineers and nine university students, asking them to complete two tasks either with or without using Gilligan.
Similarly, \citet{procrustes} also performed a user study with four software engineers and four graduate students, where each participant used Procrustes for one task and manually performed another task.

\citet{prodScalpel} conducted a user study with 20 students whose experience ranged from one to 10 years. 
Participants were divided into two groups and tasked with manually extracting and transplanting target features into the product base.

Even though there were no reports of widespread adoption of these proposed techniques in the industrial settings, the existence of user studies demonstrates the feasibility of the techniques. 
Consequently, these studies' results suggest that the techniques hold significant potential for real-world application in the future.

\begin{mdframed}
\textbf{Answer to RQ7: Our literature revealed no existing approaches to automated software transplantation to have been adopted in the industry. 
Nevertheless, \numberOfStudyThatInvolveHumanParticipants{} user studies have been conducted, \percentageOfProfessionalSoftwareEngineersInUserStudy{} of which involved SE professionals showing feasability of the approaches in practice.
Each, however, focused on a single transplantation approach, leaving the question which would be best in practice still open.
}
\end{mdframed}

\subsection{RQ8: Availability and Runnability of Software Transplantation Tools}
Beyond examining the techniques employed by different approaches and their adoption in industry, we also investigate the availability and runnability of software transplantation tools. 

In this research question, we count tools by their identity rather than by the number of publications. When multiple publications describe the same tool, we treat them as a single artifact. For example, \citet{reusing_web_application_user-interface_controls}, \citet{towards_automatic_client-side_feature_reuse} and \cite{DBLP:conf/kbse/MarasSC14} correspond to the FireCrow repository\footnote{\url{https://github.com/jomaras/Firecrow}}. In total, we identified \numberOfDistinctSoftwareTransplantationTools{} distinct tools out of the \totalNumberOfDistinctTransplantationApproaches{} approaches surveyed (see Table \ref{tab:software_transplantation_tool_availability_and_runnability}).

\begin{table}
    \centering
    \caption{Overview of \numberOfDistinctSoftwareTransplantationTools{} software transplantation tools, indicating their availability and runnability. For availability, \cmark{} denotes available for download, \xmark{} denotes unavailable, and \qmark{} denotes no response as of this writing. For runnability, \cmark{} denotes fully runnable, \tilemark{} denotes partially runnable, \ts{\dollarmark{} denotes not verified because execution requires access to a paid third-party service,} \xmark{} denotes no longer operational, and \qmark{} denotes no response as of this writing.}
    \label{tab:software_transplantation_tool_availability_and_runnability}
    \begin{tabular}{p{9cm}cc}
        \hline
        Tool Name & Availability & Runnability \\
        \hline
        AdaptivePaste~\cite{adaptivePaste} & \xmark{} & N/A \\
        AppTestMigrator~\cite{test_migration_between_mobile_apps_with_similar_functionality,automated_test_migration_for_mobile_apps} & \cmark{}\footnotemark[16] & \qmark{}\\
        CodeCarbonCopy~\cite{code_carbon_copy} & \xmark{} & N/A \\
        Code Conjurer~\cite{code_conjurer} & \cmark{}\footnotemark[17] & \xmark{} \\
        CraftDroid~\cite{test_transfer_across_mobile_apps_through_semantic_mapping} & \cmark{}\footnotemark[18] & \cmark{} \\
        FireCrow \citet{reusing_web_application_user-interface_controls, towards_automatic_client-side_feature_reuse, towards_specifying_pragmatic_software_reuse, DBLP:phd/se/Maras14, DBLP:conf/kbse/MarasSC14} & \cmark{}\footnotemark[19] & \tilemark{}\\
        Gilligan / Procrustes~\cite{gilligan,procrustes,pragmatic_software_reuse_PhD,semi_automating_pragmatic_reuse_tasks} & \ts{\xmark{}}& N/A\\
        Grow and Graft~\cite{babel_pidgin} & \xmark{} & N/A \\
        Grow and Serve~\cite{grow_django_citation} & \xmark{} & N/A \\
        GUITestMigrator~\cite{test_migration_for_efficient_large-scale_assessment_of_mobile_app_coding_assignments} & \cmark{}\footnotemark[20] & \qmark{}\\
        Hunter~\cite{hunter, hunter_fullPaper} & \xmark{} & N/A \\
        Jigsaw~\cite{jigsaw_fullPaper,jigsaw_2pages} & \qmark{} & N/A \\
        \ts{LLMigrate~\cite{automated_test_transfer_across_android_apps_using_large_language_models}} & \ts{\cmark{}\footnotemark[21]} & \ts{\dollarmark{}}\\
        NLP2TestableCode~\cite{NLP2TestableCode} & \cmark{}\footnotemark[22] & \cmark{} \\
        ProdScalpel~\cite{prodScalpel} & \cmark{}\footnotemark[23] & \qmark{} \\
        Program Fracture and Recombination~\cite{horizontal_code_transfer_via_program_fracture_and_recombination,program_fracture_and_recombination_for_efficient_automatic_code_reuse} & \xmark{} & N/A \\    
        SMALL-MINCE~\cite{test_transplantation_through_dynamic_test_slicing} & \cmark{}\footnotemark[24] & \cmark{} \\
        SmartPaste~\cite{smartPaste} & \xmark{} & N/A \\
        Splicer~\cite{program_splicing} & \qmark{} & N/A \\
        TestMig~\cite{testmig_migrating_gui_test_cases_from_ios_to_android} & \qmark{} & N/A \\
        \ts{TransDroid~\cite{gui_test_transfer_from_web_to_android}} & \ts{\cmark{}\footnotemark[25]} & \ts{\tilemark{}}\\
        \(\mu Scalpel\)~\cite{muScalpel,muScalpel_kate_ide} & \cmark{}\footnotemark[26] & \cmark{} \\
        \citet{syntax_preserving_program_slicing} & \cmark{}\footnotemark[27] & \cmark{} \\
        \citet{justyna_specialise_downstream, justyna_specialise_minisat} & \cmark{}\footnotemark[28] & \cmark{} \\
        \citet{a_study_on_code_transplantation_technique} & \qmark{} & N/A \\
        \citet{a_study_on_tree-based_code_extraction} & \qmark{} & N/A \\
        \hline
    \end{tabular}
\end{table}

\footnotetext[16]{\url{https://sites.google.com/view/apptestmigrator/}}
\footnotetext[17]{\url{https://github.com/werner-j/codeconjurer}}
\footnotetext[18]{\url{https://sites.google.com/view/craftdroid/}}
\footnotetext[19]{\url{https://github.com/jomaras/Firecrow}}
\footnotetext[20]{\url{https://sites.google.com/view/testmigration/}}
\footnotetext[21]{\url{https://github.com/seal-hub/llmigrate}}
\footnotetext[22]{\url{https://github.com/Brittany-Reid/nlp2testablecode}}
\footnotetext[23]{\url{https://autotransplantation-spl.github.io/foundry.github.io/}}
\footnotetext[24]{\url{https://github.com/mabdi/small-mince}}
\footnotetext[25]{\url{https://sites.google.com/view/icst22-transdroid}}
\footnotetext[26]{\url{http://crest.cs.ucl.ac.uk/autotransplantation/MuScalpel.html}}
\footnotetext[27]{\url{https://zenodo.org/records/7565855}}
\footnotetext[28]{\url{http://crest.cs.ucl.ac.uk/sat/satimagegi.tar.gz}}

Initially, we attempted to download all identified tools using the information provided in the respective publications. Where download links were missing or outdated, we searched public repositories (e.g., GitHub) and contacted the authors directly to request access. Based on this process, we classified the tools, in terms of availability, into three categories:
\begin{itemize}
    \item \textit{Available} — tools successfully obtained either from the publications or via author correspondence,
    \item \textit{Unavailable} — tools that authors confirmed were no longer accessible, and
    \item \textit{No response} — tools for which we contacted authors but had not received a reply as of this writing.
\end{itemize}

\begin{figure}[h]
    \centering
    \includegraphics[width=0.6\linewidth]{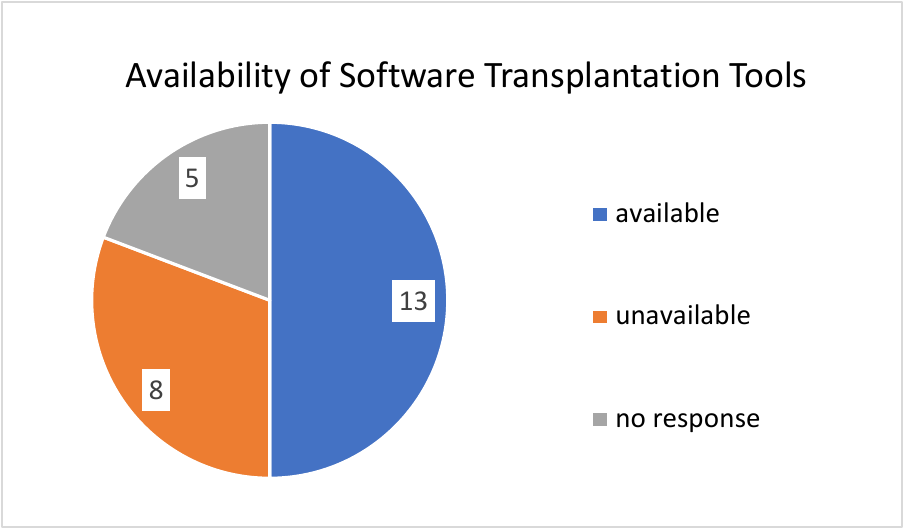}
    \caption{Availability status of the identified software transplantation tools: \percentageOfToolsAvailable{} available for download, \percentageOfToolsUnavailable{} no longer available, and \percentageOfToolsPendingResponse{} received no response from the authors as of this writing.}
    \Description{}
    \label{fig:tools_availability}
\end{figure}

Figure~\ref{fig:tools_availability} summarises the availability status. We found that \percentageOfToolsAvailable{} of tools are available for download, \percentageOfToolsUnavailable{} are no longer available, and \percentageOfToolsPendingResponse{} received no response from the authors.

Among the \numberOfToolsAvailable{} available tools, we attempted to run each one, where we prioritised the latest version of the tools. We also contacted the respective authors for clarification or support when installation or execution issues arose.

Our evaluation revealed that \numberOfAvailableToolsFullyRunnable{} tools were fully runnable without issues, including CraftDroid~\cite{test_transfer_across_mobile_apps_through_semantic_mapping}, NLP2TestableCode~\cite{NLP2TestableCode}, \(\mu\)Scalpel~\cite{muScalpel,muScalpel_kate_ide}, \citet{test_transplantation_through_dynamic_test_slicing}, \citet{syntax_preserving_program_slicing}, and \citet{justyna_specialise_downstream, justyna_specialise_minisat}.  

Only \numberOfAvailableToolsPartiallyRunnable{} tools \ts{were} partially runnable. Specifically, with FireCrow~\cite{reusing_web_application_user-interface_controls,towards_automatic_client-side_feature_reuse,DBLP:conf/kbse/MarasSC14}, we were able to execute only the organ extraction feature using recorded feature interactions. \ts{For TransDroid~\cite{gui_test_transfer_from_web_to_android}, although we were able to execute the test-transfer pipeline, the run did not produce a valid transferred test in our GitLab setup. The tool repeatedly failed to identify matching target events, resulting in \textit{EMPTY} events with a similarity score of 0 and an empty saved output. Consequently, we labeled TransDroid as partially runnable.} In contrast, \numberOfAvailableToolsNoLongerOperational{} tool, Code Conjurer~\cite{code_conjurer}, is no longer operational because its Merobase search engine backend has been discontinued.

\ts{On the other hand, we did not verify the runnability of \numberOfAvailableToolsNotVerified{} available tool. This tool is LLMigrate, whose execution depends on access to a paid third-party LLM service.}

Finally, \numberOfAvailableToolsPendingResponse{} tools require further clarification from the authors regarding their runnability. For example, we encountered runtime issues with AppTestMigrator~\cite{automated_test_migration_for_mobile_apps,test_migration_between_mobile_apps_with_similar_functionality} and GUITestMigrator~\cite{test_migration_for_efficient_large-scale_assessment_of_mobile_app_coding_assignments}, for which we received no technical feedback from the authors. Similarly, ProdScalpel~\cite{prodScalpel} could not be executed because essential files were missing from its web page. Although we received an explanation from the author, \jp{they have yet to provide the missing files as of this writing}.

\begin{mdframed}
\textbf{Answer to RQ8: We found \numberOfDistinctSoftwareTransplantationTools{} distinct software transplantation tools in the literature. 
Of these, \percentageOfToolsAvailable{} are available for download, \percentageOfToolsUnavailable{} are no longer available.
Availability of others is unclear.
Among the available tools, \numberOfAvailableToolsFullyRunnable{} are fully runnable, \numberOfAvailableToolsPartiallyRunnable{} is partially runnable, and \numberOfAvailableToolsNoLongerOperational{} is no longer operational.}
\end{mdframed}

\section{Software Transplantation Challenges}
In this section, we provide a comprehensive discussion of our findings. 
We first compare and contrast different approaches to software transplantation. 
Then, we critically analyse the experimental designs as well as the existing benchmarks used in the literature.  
Moreover, we identify the bottlenecks to adopting software transplantation in an industrial setting. 
We provide recommendations for potential solutions to each issue discussed.

\subsection{Software Transplantation Approach}
This section highlights and discusses the similarities and differences between the proposed approaches. 
We analyse and compare the proposed techniques based on the \(LATIIV^R\) framework~\cite{genetic_programming_for_reverse_engineering}, as discussed in Section~\ref{sec:software_transplantation_history}.

\subsubsection{Localise} 
From RQ2, we observed that nearly all software transplantation techniques do not automatically locate an organ. Instead, they either require software engineers to provide an organ, which can be in various formats (e.g., organ entry point or reuse plan), or ask for a search query that will later be used to find an organ.
However, \numberOfApproachesNotRequireOrganLocation{} approaches do not require the organ to be located.
For example, \citet{babel_pidgin} and \citet{grow_django_citation} automated the discovery of donor logic, using fitness functions to identify and extract functional organs from within a host's existing API surface.
\citet{horizontal_code_transfer_via_program_fracture_and_recombination, program_fracture_and_recombination_for_efficient_automatic_code_reuse} automatically selects a target shard for the task.
\citet{justyna_specialise_downstream, justyna_specialise_minisat} uses `code bank' as an organ.

\subsubsection{Target}
The process of determining the location for the organ to be transplanted into the host remains largely manual.
\numberOfApproachesUserProvidesInsertionLocation{} approaches (\percentageOfApproachesUserProvidesInsertionLocation{}) still require software engineers to specify this location.
For example, \(\mu Scalpel\) \cite{muScalpel} asks users to label the insertion site, while in the code search domain, NLP2TestableCode \cite{NLP2TestableCode} requests users to provide a search query for the line where the source code should be inserted.
For Grow and Graft~\cite{babel_pidgin}, users provide a list of possible insertion locations.

In contrast, \numberOfApproachesDoNotMentionInsertionLocation{} approaches (\percentageOfApproachesDoNotMentionInsertionLocation{}) did not mention any detail about the insertion location, such as \citet{a_study_on_tree-based_code_extraction}, \citet{a_study_on_code_transplantation_technique}, \citet{test_transplantation_through_dynamic_test_slicing}, Skipper~\cite{validating_pragmatic_reuse_tasks}, and Program Fracture and Recombination~\cite{horizontal_code_transfer_via_program_fracture_and_recombination,program_fracture_and_recombination_for_efficient_automatic_code_reuse}.

Lastly, \numberOfApproachesRandomlyInsertOrgan{} approach (\percentageOfApproachesRandomlyInsertOrgan{}) inserts the organ in a random location \citet{justyna_specialise_downstream, justyna_specialise_minisat}.

\subsubsection{Interface and Insert}
To transplant an organ into the host, many methods were proposed, which rely on different assumptions and are capable of supporting different levels of transplantation. 

First, \numberOfApproachesUsingGPForOrganAdaptation{} approaches (\percentageOfApproachesUsingGPForOrganAdaptation{}) rely on a heuristic algorithm (e.g., genetic programming) to adapt an organ to the host's environment. 
Such techniques rely on testing to check the validity of the transplant. 
These include \(\mu Scalpel\) \cite{muScalpel, muScalpel_kate_ide}, ProdScalpel \cite{prodScalpel}, software specialisation \cite{justyna_specialise_downstream,justyna_specialise_minisat}, Grow and Graft \cite{babel_pidgin} and Grow and Serve \cite{grow_django_citation}.

Second, the adaptation takes place after the organ is implanted into the host for \numberOfApproachesAdaptingOrganAfterInsertion{} approaches (\percentageOfApproachesAdaptingOrganAfterInsertion{}).
These include pragmatic software reuse \cite{gilligan, procrustes}, CodeConjurer \cite{code_conjurer}, NLP2TestableCode \cite{NLP2TestableCode}, Splicer \cite{program_splicing}, AdaptivePaste\cite{adaptivePaste}, and SmartPaste \cite{smartPaste}.

Third, \numberOfApproachesInsertingOnlyMissingPartOfOrganIntoHost{} approaches (\percentageOfApproachesInsertingOnlyMissingPartOfOrganIntoHost{}) compare the differences between the organ and the host and only insert the missing parts into the host. 
For instance, Jigsaw~\cite{jigsaw_2pages, jigsaw_fullPaper} and~\citet{syntax_preserving_program_slicing} are the only approaches that transplant specific parts of the code that are missing. 
While Jigsaw achieves this by doing a node mapping in their augmented version of AST (i.e., CAST), \citet{syntax_preserving_program_slicing} performs a string comparison between an organ and its target host in the format of markup code.

Similarly, \(\mu Scalpel\) \cite{muScalpel} only transplants as many statements as necessary to make an organ work in the host. 
\(\mu Scalpel\) relies on the assumption that \textit{``our organ needs very few of the statements in their donor''}~\cite{muScalpel}.

Fourth, we found that \numberOfApproachesEnumeratePossibleInsertionLocations{} technique (\percentageOfApproachesEnumeratePossibleInsertionLocations{}) simply enumerates over all possible locations for implantation as a means of transplantation. 
\citet{babel_pidgin} grow a language translation functionality and enumerate all possible locations to find the place for inserting the grown functionality.

\subsubsection{Validation} 
In RQ4, we observed that \percentageOfApproachesUsingInternalValidationStep{} of the software transplantation approaches have an automated validation step integrated within the approach.
\ts{Among those with integrated validation steps, \numberOfApproachesUsingInternalValidationStepViaTest{} out of \numberOfApproachesUsingInternalValidationStep{} approaches relies on test cases to confirm that the new functionality has been added to the host and/or to check for regression faults.}
In contrast, there are \numberOfApproachesUsingInternalValidationStepViaOtherThanTest{} approaches that use validation techniques other than test cases.
For example, Splicer \cite{program_splicing} allows users to provide either test cases or finite automata for validation.
Conversely, \citet{program_fracture_and_recombination_for_efficient_automatic_code_reuse, horizontal_code_transfer_via_program_fracture_and_recombination} rely on shard matching (e.g., input and output) to find the correct organ for the host.

\subsection{Experimental Design} \label{sec:discussion_experimental_design}
This section discusses the experimental design of the software transplantation approaches.
See Section \ref{sec:rq5} for more detail about the validation technique employed by each transplantation approach.

Our literature review shows that while most software transplantation studies include experiments or proof-of-concept implementation to test their proposed techniques, they rarely compare them with existing approaches.
\ts{Only AppTestMigrator~\cite{test_migration_between_mobile_apps_with_similar_functionality}, LLMigrate~\cite{automated_test_transfer_across_android_apps_using_large_language_models}, and Splicer~\cite{program_splicing} perform such a comparison.}

While such comparisons are valuable for demonstrating improvements or identifying limitations, they are not always feasible.
This is because differences in approach's assumptions and implementation availability can make direct comparisons challenging, such as machine learning for renaming variable~\cite{smartPaste} versus adding missing lines and files to obtain new functionality~\cite{syntax_preserving_program_slicing}.
Nonetheless, comparison against other approaches within the same domain and under similar assumptions remains important, as it helps clarify the contribution of new techniques and avoids the risk of simply reinventing the wheel.

Moreover, we found that only \numberOfStudyThatInvolveHumanParticipants{} studies conducted a user study to validate their proposed technique, indicating that user studies are not commonly used in the software transplantation domain.
Among these, only \percentageOfProfessionalSoftwareEngineersInUserStudy{} of the participants in the user study were professional software engineers.
Although user studies provide valuable insights into the usability of proposed approaches, they often lack participation from individuals who represent the intended user base.
This is because university students may not yet be fully competent in terms of knowledge and skills in professional software development workflow and practice.

The difficulty in recruiting professional software engineers for user studies is a significant challenge.
Practitioners are typically engaged in time-sensitive development work, making them less available for research studies.
Additionally, participation may require employer approval, add scheduling complexity, and involve legal or confidentiality concerns.
This is particularly true when industrial use cases or proprietary systems are involved.

\subsubsection{Recommendations}

We recommend user studies within the industry to establish the true potential of the proposed transplantation approaches in a real-world setting.
Researchers could consider offering incentives to encourage participation, such as providing free training or workshops on the proposed approach.
This could help to increase professional software engineers participation.
Such efforts not only strengthen the evaluation of a technique but can also serve as a bridge to industry adoption by demonstrating practical value and usability in authentic development environments.

\subsection{Benchmarks} \label{sec:discussion_benchmarks}
This section discusses the benchmarks commonly used in the literature to evaluate proposed software transplantation techniques.

\ts{For this analysis, we combine the benchmark availability categories into three broader classes (see Section \ref{sec:rq_benchmarking_empirical_benchmark}). The broader classes are as follows:
\begin{itemize}
\item \textit{Available} - benchmarks that are open source,
\item \textit{Conditionally available} - benchmarks for which no licence information is provided or whose retrieval requires tool execution,
\item \textit{Unavailable} - benchmarks for which no artefact link is provided, benchmarks whose source code is unavailable, and artificial benchmarks that were not released as reusable artefacts.
\end{itemize}
}

\ts{From our literature review, we found that source code of \numberOfAvailableBenchmarksTotal{} benchmarks (\percentageOfAvailableBenchmarksTotal{}) are available, \numberOfConditionallyAvailableBenchmarksTotal{} benchmarks (\percentageOfConditionallyAvailableBenchmarksTotal{}) are conditionally available, and \numberOfUnavailableBenchmarksTotal{}  benchmarks (\percentageOfUnavailableBenchmarksTotal{}) are unavailable.}

\begin{figure}[]
    \centering
    \includegraphics[width=0.6\linewidth]{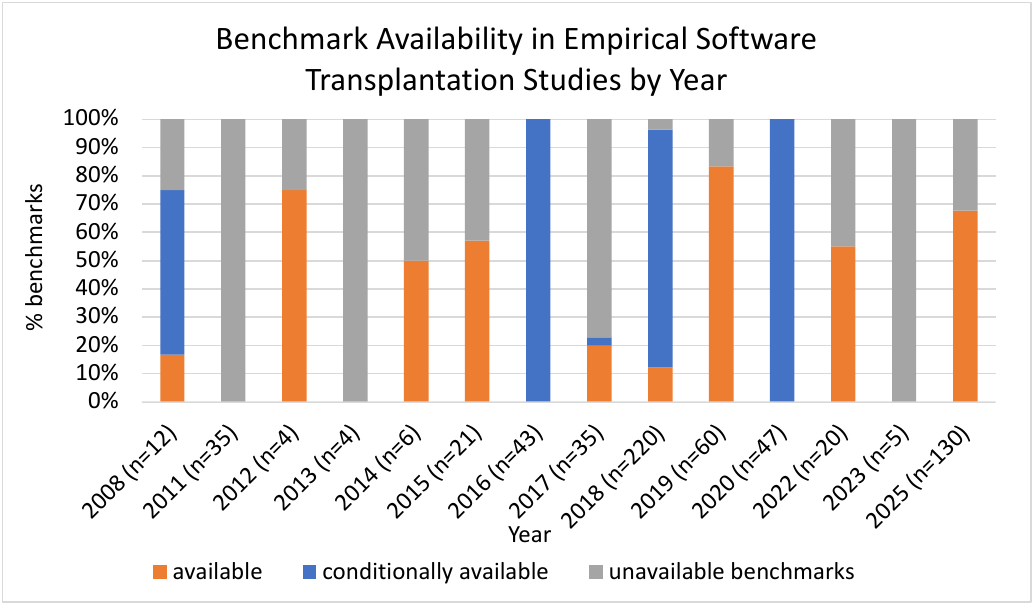}
    \caption{\ts{Benchmark availability in empirical software transplantation studies by year. Each bar is normalised to 100\% and shows the proportion of benchmarks classified as available, conditionally available, or unavailable. The number of benchmark instances considered in each year is shown in parentheses.}}
    \label{fig:benchmarks_availability_by_year}
    \Description{}
\end{figure}

\ts{Figure \ref{fig:benchmarks_availability_by_year} shows the availability of software artefacts from 2008 to 2025.
While some years show a high degree of unavailablility, with some (e.g., 2011, 2013, and 2023) reaching 100\%, some years contain a larger proportion of available and conditionally available benchmarks, particularly in 2016, 2018, 2019, and 2020.
Overall, the data does not suggest a steady improvement in benchmark availability over time.
These findings raise significant concerns about the reproducibility of experiments, as rerunning studies with the same benchmarks is often impossible.
Moreover, this gap hinders researchers' ability to compare new approaches with existing ones, limiting progress in the field.}

This issue is similar to the one discussed in Section \ref{sec:discussion_experimental_design}, where we noted that comparisons with existing approaches are also rare.
In both cases, a lack of publicly available benchmarks makes it difficult to conduct fair and meaningful comparisons --- whether across techniques or through user evaluations.
This further emphasizes the need for community efforts toward building shared datasets and evaluation criteria.

Furthermore, we observed that software transplantation techniques have not been frequently applied to modern benchmarks or programming languages commonly used to develop the software we interact with on a daily basis.
In today's world, people heavily rely on software offered through various online platforms, which often depend on cloud and web services.
These services are typically built using modern and high-level languages such as the Go Programming Language (i.e., Golang) and JavaScript.
This gap in research raises important questions about whether existing transplantation techniques are suitable for cloud-based and web development environments.
Empirical studies going beyond Java/C might gather more interest within the industry and increase uptake of transplantation techniques.

\subsubsection{Recommendations}
In order to foster research in automated software transplantation, we recommend creation of a targeted, publicly available benchmark on which multiple transplantation approaches could be compared against. 
We recommend the benchmark provides donor, hosts, and organs for different application domains, such as web or cloud based services, to explore how different software architectures influences the effectiveness of software transplantation.

\subsection{Synthetic Alternatives: LLMs and the Maturity Gap}
In this section we discuss why LLMs cannot {\em replace} software transplantation and how can they help overcome remaining barriers to widespread adoption and deployment of the transplantation techniques.

By definition, newly generated code by LLMs is not `battle tested'. 
It has not been deployed over many years, and thereby undergone the rigorous beta-testing  testing that takes place in production.
\ts{By contrast, existing code that has been deployed for several years, whether written by language model, human or some hybrid of the two, has this hugely important property of being `battle tested'.}
As such, it is known to be fit for purpose; a guarantee that can only come from deployment in production.
It is aphorism  that ``no plan survives contact with the enemy''\footnote{This quote is a paraphrase of the original, in German, that has been attributed to Helmuth von Moltke the Elder (1800-1891).}.
We might say the same thing of untested code that has never seen the figurative `battles' that code goes through in production; 
it is inherently unknown whether it will survive contact with real users.

\subsubsection{Recommendations}
Although LLMs cannot, by definition, construct battle tested code,
they have the potential to assist various stages of the software transplantation pipeline and help address key bottlenecks in adopting automated transplantation techniques. 
For example, \citet{instruct_or_interact_exploring_and_eliciting_llms_capability_in_code_snippet_adaptation_through_prompt_engineering} investigate the use of LLMs for automated code snippet adaptation to new contexts, highlighting their potential applicability to transplantation-related tasks. Similarly, \citet{togll_correct_and_strong_test_oracle_generation_with_llms} explore the use of LLMs to generate test oracles and demonstrate that LLM-generated assertions can improve fault detection when paired with automatically generated test cases. However, they also highlight limitations related to robustness and sensitivity to prompt design. 
Nevertheless, such approaches typically address isolated tasks and are not framed as end-to-end transplantation techniques. They often rely on interactive or human-in-the-loop processes. Consequently, LLMs are best viewed as complementary tools that can support individual stages of the software transplantation pipeline rather than as standalone solutions for automated software transplantation.
We expected the advent of language models and agentic software development will remove the remaining barriers 
to widespread adoption and deployment of automated software transplantation.
LLMs have the potential to automate the remaining manual phases identified in this survey which are the primary remaining barriers to adoption.

\subsection{Mechanistic Bottlenecks: Donor Identification and Grafting}
In this section, we identify the potential bottlenecks that we believe may be the primary reasons preventing industrial practitioners from adopting software transplantation, despite the promising experimental results.

We hypothesize that a primary reason industrial practitioners have not adopted automated software transplantation is the substantial manual effort required before the transplantation process can begin. Based on the existing literature, we identify two key manual pre-requisite steps.

First, software engineers have to manually locate the organ, i.e., feature of interest, within the donor system. 
Current techniques typically require the engineer to either specify an organ entry point or provide a search query. 
When specifying an organ entry point, the engineer needs to label the start with a special comment.
The process of searching for the start of the organ heavily depends on the total number of files or Lines of Code (i.e., LOC) that the software has. 
In the case of legacy software, where there is no documentation, searching for specific features of interest might not even be feasible.

Similarly, some approaches attempt to assist organ identification through code search by requiring an engineer to provide a query describing the desired functionality. 
While this reduces manual navigation, it still relies on the \ts{engineer's} ability to formulate an effective query and has so far only been demonstrated at small scale within controlled code search settings. 
As a result, even these semi-automated solutions do not eliminate the manual pre-requisite effort required to initiate transplantation.

Second, software engineers often need to manually augment or create a test suite for the organ to validate its functionality after transplantation. 
Augmentation requires an engineer to thoroughly examine the donor's test suites and rewrite them for the host, which includes the translation of test oracle from donor's into host's context. 
Yet, different coding styles and conventions between host and donor software could pose a further challenge as the interfaces are different, and existing test suites could not be mapped directly. 
Furthermore, existing host's tests might need expanding to catch regressions. 
This adds extra workload on software engineers.

\subsubsection{Recommendations}
To reduce the manual effort associated with organ location, we recommend leveraging code search techniques at the repository level. 
Instead of searching a database of code snippets, code search engines can be applied directly to the donor repository to identify candidate entry points for an organ. 
This shifts the developer's effort from manually inspecting numerous files to providing a structured query, which may take the form of a natural-language description, function prototype, or class interface. 
While this approach does not fully eliminate manual input, it can substantially reduce the cost of initiating the transplantation process.

Alternatively, the donor's existing test suite can be leveraged to assist organ location. 
By executing the tests associated with the target organ, only the program statements exercised by those tests will be executed. 
Consequently, this provides an execution-based approximation of the organ's location within the repository. 
To further refine this approximation, similarity metrics (e.g., Jaccard index) can be used to measure the overlap between execution traces of different test cases and identify source code elements that are consistently associated with the target functionality.

A limitation of this approach is that individual test cases often exercise only a subset of the feature's behaviour and may not provide complete coverage of the organ. 
Consequently, a sufficiently diverse and comprehensive set of test cases is required to reliably capture the full extent of the organ, which may not always be available in practice.

With the recent rise of LLMs, we believe that there is potential in using LLMs to locate a feature of interest. 
Given that LLM relies on natural language as an input, we could supply it with the donor's document or our own description of the target feature and ask LLMs to locate the start of an organ from the donor's repository.

However, the use of LLMs does not eliminate the need for manual input. Engineers must still clearly and efficiently articulate their intent by describing the desired behaviour, constraints, and context of the feature they wish to transplant. Poorly specified or ambiguous descriptions may lead to imprecise or irrelevant results, particularly for large or complex codebases. While LLMs reduce the effort of navigating and understanding unfamiliar code, they shift the manual burden from code navigation to intent specification.

Regarding use of testing to validate transplantations, currently engineers need to manually either create or augment existing tests. 
We propose leveraging LLMs to reduce this manual effort.
One approach is to provide the LLM with documentation or textual descriptions of the feature of interest and ask it to generate corresponding test cases. 
Alternatively, we could extract a program slice that contains the target feature and use the LLM to generate a test suite specifically for that code fragment.

Although this idea has not been directly applied in the software transplantation domain, many publications demonstrate the feasibility of the idea in similar use cases. 
For instance, \citet{searchgem5_towards_reliable_gem5_with_search_based_software_testing} proposed an idea of using LLM to parameterise the C program to generate the test cases for Gem5 software. 
\citet{how_well_does_llm_generate_security_test} demonstrated that LLM can be used to generate the security test suites to analyse the vulnerabilities of third-party libraries. 
These studies indicate that LLMs could potentially automate the creation or augmentation of test suites in transplantation scenarios, reducing the need for manual rewriting or adaptation of tests.

\section{Related Work}
This section reviews related work and explores the applications of software transplantation in other domains.

\paragraph{Other Related Surveys}
An early overview of research related to software transplantation is provided by \citet{an_insight_on_software_features_supporting_software_transplantation}. This study surveys a subset of publications with the aim of broadly characterising research directions connected to software features and reuse. Rather than offering an in-depth or systematic analysis of software transplantation techniques, the review is exploratory in nature and primarily serves as a scoping exercise to introduce readers to neighbouring research areas.

In contrast, our work presents a systematic literature review that focuses specifically on software transplantation as a distinct research domain. Following established systematic review guidelines~\cite{guideline_for_performing_systematic_literature_review}, we employ domain-specific search terms (e.g., ``software transplantation'' and ``code transplantation'') to identify relevant studies and ensure methodological rigour and reproducibility. This allows us to provide a more comprehensive view of the field by examining how transplantation approaches are evaluated and by identifying limitations that may help explain the lack of widespread industrial adoption. Further details of our review methodology are presented in Section~\ref{sec:survey_method}. Readers seeking a broad introductory overview of feature-related research areas may refer to the earlier review by \citet{an_insight_on_software_features_supporting_software_transplantation}.

\paragraph{Copy-and-Paste}
Reusing an existing piece of code during software development is a common practice~\cite{an_empirical_study_of_the_copy_and_paste_behaviour_during_development, how_do_developers_adapt_code_snippets_to_their_contexts_an_empirical_study_of_context_based_code_snippet_adaptations}. 
This can be realised via a form of abstraction at various levels, such as functions or classes. 
However, designing appropriate abstractions often requires significant time and effort. 
Alternatively, a software engineer may resort to a `copy-paste' approach, where an existing piece of code is copied to the new location and adapted to fit the new context. 
Such practice avoid the risk of unintentionally altering the existing system~\cite{cloning_considered_harmful_considered_harmful_patterns_of_cloning_in_software} and can be significant faster than manual abstraction~\cite{copy_and_paste_redeemed}, making it popular among software engineers.

Despite its convenience, copy-and-paste reuse introduces several drawbacks. 
Code cloning can increase software complexity, propagate bugs, and complicate long-term maintenance~\cite{copy_and_paste_redeemed}. 
The adaptation process remains largely manual and ad hoc. 
In contrast, automated software transplantation aims to support this reuse process by automating each step of extraction, adaptation and integration. 
It specialises the organ to the host's environment aiming to get rid of unneccessary bloat.

\paragraph{Software Product Line}
A Software Product Line (SPL) is a family of software systems that share a set of common features but differ in some specific features ~\cite{software_product_line_engineering_and_variability_management}.  Software features lie at the core of SPL. These are the core assets that are reused in different products~\cite{software_product_lines}. A key benefit of SPL is systematic reuse, which enables company to reduce development effort and time-to-market by deriving multiple products from a shared set of core assets rather than developing each product independently.

Although SPL promises a great benefit, it has been shown to introduce multiple challenges, including high upfront investment in core asset and variability design before products can be derived~\cite{software_product_line_evaluation_in_the_large}, difficulty in managing variability across a product line~\cite{performance_variability_in_software_product_lines_proposing_theories_from_a_case_study}, and substantial testing overhead owing to the large number of possible product configurations~\cite{on_strategies_for_testing_software_product_lines_a_systematic_literature_review}. In contrast, automated software transplantation enables opportunistic reuse of functionality across existing systems without requiring a predefined variability model or centralised platform. This offers a more lightweight alternative in scenarios where the costs and rigidity of SPLE are prohibitive.

\paragraph{Refactoring}
The term ``refactoring'' was first introduced by \citet{refactoring_a_program_restructuring_aid_in_designing_object_oriented_application_frameworks} in his PhD dissertation, who defined it as a process of restructuring the existing code to improve its internal structure while preserving the external behaviour of the code. As software evolves to accommodate changing user requirements, code complexity tends to increase, which can lead to the degradation of internal quality attributes such as maintainability. Refactoring addresses this issue by restoring internal software quality without altering software behaviour.

Nevertheless, refactoring differs from automated software transplantation in two key aspects. First, refactoring does not introduce new functionality into a software system. Second, it operates within a single software system. In contrast, automated software transplantation focuses on transferring and integrating functionality across different software systems, whereas refactoring aims solely to improve internal structure while preserving existing behaviour.

\paragraph{LLMs {\em Instead of} Software Transplantation}

With the recent rise of Large Language Models (LLMs), 
what might think that we do not need a {\em specific} approach to transplantation; why not just use language models to generate code as needed?
LLMs provide new opportunities for software generation through generative interfaces trained on vast corpora of source code. 
These models can produce reusable code from natural language prompts, enabling developers to rapidly obtain candidate implementations. 
However, LLMs do not fully resolve the challenges of trusted software reuse. 
Generated outputs are often criticised for being inefficient~\citep{effi_code_jie_zhang} 
or verbose~\citep{whats_wrong_with_your_code_generated_by_large_language_models}. 
Moreover, LLM-generated code lacks strong correctness guarantees and may appear plausible while still containing subtle bugs~\citep{bugs_in_llms_tambon, ticoder_subtle_bugs}.

\section{Threats To Validity}
This section outlines potential threats to validity and the steps we took to address them.

One author was responsible for identifying relevant literature.
To reduce the risk of missing important publications, we uniformly sampled 10\% of the papers (N = 159 papers) and asked a second author to independently evaluate their relevance based on the scope of this literature review.
The two authors initially disagreed on only four papers. 
These disagreements were resolved through discussion, resulting in a shared consensus. 
Furthermore, we reached out to authors of identified relevant papers to further ensure we did not miss relevant work. 
This resulted in additional 4 papers added to our literature review.

Regarding the selection of keywords, we exercised caution to ensure we used the most relevant technical terms for the search.
As described in Section \ref{sec:survey_method}, we initially performed a primary search using the keyword `software transplantation'.
However, we discovered that some publications use the term `code transplantation' to describe the same domain.
As a result, we used the two terms and their grammatical variations, i.e., software transplant* and code transplant*'.
Additionally, we explored alternative keywords like `feature migration' and `code migration'. 
However, we found that the results were not relevant to our study, as did not cover the whole transplantation pipeline, from organ location through its implantation. 
Therefore, these terms were excluded from our final keyword list.

\section{Conclusions}

Automated software transplantation has been introduced as a paradigm to remove the labour-intensive task of transferring and adapting functionality, i.e, an organ, from a donor codebase to the target, host system. 
Here, we present the first literature review on this domain to understand the current state-of-the-art and identify opportunities  for its wider adoption.

We found that the most frequently reused artefact is a software feature (in \percentageOfFeatureReuseApproaches{} of reported approaches), followed by a test case (in \percentageOfTestReuseApproaches{} of the reported approaches). 
In terms of organ extraction, \percentageOfApproachesUsingProgramSlicingForOrganExtraction{} of software transplantation approaches rely on program slicing and another \percentageOfApproachesUsingCodeSearchEngineForOrganExtraction{} rely on a code search engine.
At the same time, genetic programming has been applied both to adapt the organ to the host environment and to generate new functionality from scratch for \percentageOfApproachesUsingGPForOrganAdaptation{} of the proposed approaches.
Only \percentageOfApproachesUsingInternalValidationStep{} of the proposed techniques have some form of validation mechanism integrated within the approach.
We found \numberOfStudyThatInvolveHumanParticipants{} articles that present a user study with professional software engineers and/or university students showing viability of the approach in a real-world setting.

Furthermore, we offer suggestions for addressing potential bottlenecks to wider industrial adoption of automated software transplantation techniques.  
These include automation of the manual steps for organ identification and test suite augmentation for the host environment for validation of the transplanted code.
Large Language Models (LLMs) have introduced a transformative paradigm for reuse by distilling the vast corpus of code within their training data into a generative interface. 
While they have revolutionised reuse, they do not completely solve the problem, as they often generate slightly incorrect, suboptimal code, and struggle with domain-specific logic.
We posit they can, however, unlock the remaining barriers to wider adoption of software transplantation.

We hope this survey will boost further research in the automated software transplantation domain.

\section{Acknowledgements}

\jp{We would like to thank the authors of the work cited here who provided feedback on an earlier draft of this survey.}

\bibliographystyle{ACM-Reference-Format}
\bibliography{Bibliography}
\end{document}